\documentclass[10pt,conference]{IEEEtran}

\usepackage[dvipsnames]{xcolor}
\usepackage{booktabs} % For formal tables
\usepackage{tabularx}
\usepackage{colortbl}
\usepackage{arydshln}
\usepackage{adjustbox}
\usepackage[linesnumbered,ruled,vlined]{algorithm2e}
\usepackage{amsmath}
 
\usepackage{amssymb}
\usepackage{bm}
\usepackage{enumitem}
\usepackage{soul}
\usepackage{threeparttable}  
\usepackage{multirow}
\usepackage{makecell}
\usepackage{tabularx}
\usepackage[T1]{fontenc}
\usepackage{array}
\usepackage{listings}
\usepackage{xspace} 
\usepackage{subcaption}
\usepackage{graphicx}
\usepackage{hyperref}
\usepackage{textcomp}
\usepackage{microtype} 
\usepackage{calc}
\usepackage{pifont}
\usepackage{tcolorbox}
\usepackage{lscape}
\usepackage{rotating}
\usepackage{balance}
\usepackage{cite}

\newcolumntype{Y}{>{\centering\arraybackslash}X}

\SetCommentSty{mycommfont}
\SetKwInput{KwInput}{Input}                % Set the Input
\SetKwInput{KwOutput}{Output}              % set the Output

\newlength{\ColorBoxDepthReference}
\newlength{\ColorBoxHeightReference}
\newlength{\Width}%
\newcommand{\MyColorBox}[2][red]%
{%
	\settowidth{\Width}{#2}%
	\colorbox{#1}%
	{%      
		\raisebox{-\ColorBoxDepthReference}%
		{%
			\parbox[b][\ColorBoxHeightReference+\ColorBoxDepthReference][c]{\Width}{\centering#2}%
		}%
	}%
}
\definecolor{codegreen}{rgb}{0,0.6,0}
\definecolor{codegray}{rgb}{0.5,0.5,0.5}
\definecolor{codepurple}{rgb}{0.58,0,0.82}
\newcommand{\inlinecode}[1]{``\texttt{#1}''}

\newcommand{\best}[1]{\cellcolor{gray!25}#1}

\newcommand{\summary}[1]{
	\begin{center}
		\begin{tcolorbox}[colback=gray!10,colframe=black!25,width=1\columnwidth,arc=1mm, auto outer arc,boxrule=0.5pt,boxsep=5pt,left=3pt,right=3pt,top=0pt,bottom=0pt]
		\textbf{SUMMARY:} #1
		\end{tcolorbox}
	\end{center}
}

\newcommand{\eg}{\textit{e.g.,}\xspace}
\newcommand{\ie}{\textit{i.e.,}\xspace}
\newcommand{\etal}{\textit{et al.}\xspace}

\newcommand{\App}{\textsc{ToolGen}\xspace}

\newcommand{\AppGPT}{\textsc{ToolGen}-\emph{gpt}\xspace}
\newcommand{\AppTFive}{\textsc{ToolGen}-\emph{t5}\xspace}

\newcommand{\AppLlama}{\textsc{ToolGen}-\emph{llama}\xspace}

\newcommand{\RAGAppGPT}{\textit{rag}\textsc{ToolGen}-\emph{gpt}\xspace}
\newcommand{\RAGAppTFive}{\textit{rag}\textsc{ToolGen}-\emph{t5}\xspace}

\newcommand{\RAGAppLlama}{\textit{rag}\textsc{ToolGen}-\emph{llama}\xspace}

\newcommand{\Base}{\textsc{Vanilla}\xspace}
\newcommand{\BaseGPT}{\textsc{Vanilla}-\emph{gpt}\xspace}
\newcommand{\BaseTFive}{\textsc{Vanilla}-\emph{t5}\xspace}

\newcommand{\BaseLlama}{\textsc{Vanilla}-\emph{llama}\xspace}

\newcommand{\Repo}{\textsc{RepoCoder}\xspace}
\newcommand{\RepoGPT}{\textsc{RepoCoder}-\emph{gpt}\xspace}
\newcommand{\RepoTFive}{\textsc{RepoCoder}-\emph{t5}\xspace}

\newcommand{\RepoLlama}{\textsc{RepoCoder}-\emph{llama}\xspace}

\newcommand{\comp}{\texttt{\small <COMP>}\xspace}
\newcommand{\bos}{\texttt{\small <BOS>}\xspace}
\newcommand{\eos}{\texttt{\small <EOS>}\xspace}

\newcommand{\undefv}{\emph{undefined-variable}\xspace}
\newcommand{\nomem}{\emph{no-member}\xspace}

\newcommand{\cov}{\emph{Dependency Coverage}\xspace}
\newcommand{\valid}{\emph{Static Validity Rate}\xspace}

\newcommand{\revise}[1]{\textcolor{black}{#1}}

\begin{document}

\title{Teaching Code LLMs to Use Autocompletion Tools in Repository-Level Code Generation}

\author{
    \IEEEauthorblockN{Chong Wang\IEEEauthorrefmark{1}, Jian Zhang\IEEEauthorrefmark{1,3}, Yebo Feng\IEEEauthorrefmark{1}, Tianlin Li\IEEEauthorrefmark{1}, Weisong Sun\IEEEauthorrefmark{1}, Yang Liu\IEEEauthorrefmark{1}, and Xin Peng\IEEEauthorrefmark{2}}
    \IEEEauthorblockA{
        \IEEEauthorrefmark{1}\textit{School of Computer Science and Engineering, Nanyang Technological University, Singapore}\\
        \{chong.wang, jian\_zhang, yebo.feng\}@ntu.edu.sg, tianlin001@e.ntu.edu.sg, \{weisong.sun, yangliu\}@ntu.edu.sg\\
        \IEEEauthorrefmark{2}\textit{School of Computer Science and Shanghai Key Laboratory of Data Science, Fudan University, China}\\ 
        pengxin@fudan.edu.cn\\
    }
    % \thanks{\IEEEauthorrefmark{3} J. Zhang is the corresponding author}
}

% \makeatletter
% \newcommand{\linebreakand}{%
%   \end{@IEEEauthorhalign}
%   \hfill\mbox{}\par
%   \mbox{}\hfill\begin{@IEEEauthorhalign}
% }
% \makeatother

% \author{\IEEEauthorblockN{Chong Wang}
% \IEEEauthorblockA{\textit{Nanyang Technological University} \\
% Singapore, Singapore \\
% chong.wang@ntu.edu.sg}
% \and
% \IEEEauthorblockN{Jian Zhang}
% \IEEEauthorblockA{\textit{Nanyang Technological University} \\
% Singapore, Singapore \\
% jian\_zhang@ntu.edu.sg}
% \and
% \IEEEauthorblockN{Yebo Feng}
% \IEEEauthorblockA{\textit{Nanyang Technological University} \\
% Singapore, Singapore \\
% yebo.feng@ntu.edu.sg}
% \linebreakand
% \IEEEauthorblockN{Tianlin Li}
% \IEEEauthorblockA{\textit{Nanyang Technological University} \\
% Singapore, Singapore \\
% tianlin001@e.ntu.edu.sg}
% \and
% \IEEEauthorblockN{Weisong Sun}
% \IEEEauthorblockA{\textit{Nanyang Technological University} \\
% Singapore, Singapore \\
% weisong.sun@ntu.edu.sg}
% \and
% \IEEEauthorblockN{Yang Liu}
% \IEEEauthorblockA{\textit{Nanyang Technological University} \\
% Singapore, Singapore \\
% yangliu@ntu.edu.sg}
% \linebreakand
% \IEEEauthorblockN{Xin Peng}
% \IEEEauthorblockA{\textit{Fudan University} \\
% Shanghai, China \\
% pengxin@fudan.edu.cn}
% }

\maketitle

\begin{abstract}
    Recent code large language models (LLMs) have shown promising performance in generating standalone functions. However, they face limitations in repository-level code generation due to their lack of awareness of \emph{repository-level dependencies} (\eg user-defined attributes), resulting in \emph{dependency errors} such as undefined-variable and no-member errors. In this work, we introduce \App, an approach that integrates autocompletion tools into the code LLM generation process to address these dependencies. \App comprises two main phases: Trigger Insertion and Model Fine-tuning (Offline), and Tool-integrated Code Generation (Online). During the offline phase, \App augments functions within a given code corpus with a special mark token, indicating positions to trigger autocompletion tools. These augmented functions, along with their corresponding descriptions, are then used to fine-tune a selected code LLM. In the online phase, \App iteratively generates functions by predicting tokens step-by-step using the fine-tuned LLM. Whenever a mark token is encountered, \App invokes the autocompletion tool to suggest code completions and selects the most appropriate one through constrained greedy search. 
    
    We conduct comprehensive experiments to evaluate \App's effectiveness in repository-level code generation across three distinct code LLMs: CodeGPT, CodeT5, and CodeLlama. To facilitate this evaluation, we create a benchmark comprising 671 real-world code repositories and introduce two new dependency-based metrics: \cov and \valid. \revise{The results demonstrate that \App significantly improves \cov by 31.4\% to 39.1\% and \valid by 44.9\% to 57.7\% across the three LLMs, while maintaining competitive or improved performance in widely recognized similarity metrics such as BLEU-4, CodeBLEU, Edit Similarity, and Exact Match.
    On the CoderEval dataset, \App achieves improvements of 40.0\% and 25.0\% in test pass rate (Pass@1) for CodeT5 and CodeLlama, respectively, while maintaining the same pass rate for CodeGPT. \App also demonstrates high efficiency in repository-level code generation, with latency ranging from 0.63 to 2.34 seconds for generating each function.} Furthermore, our generalizability evaluation confirms \App's consistent performance when applied to diverse code LLMs, encompassing various model architectures and scales.
\end{abstract}
\maketitle

\section{Introduction}\label{sec:intro}

% \textbf{Background.}
Code generation has been a longstanding focal point in the field of software engineering. Recent advancements have introduced a variety of code large language models (LLMs)~\cite{gpt-neo,gpt-j,Codex,CodeT5,CodeT5Plus,InCoder,AlphaCode,CodeGen,GPT-NeoX,SantaCoder,StarCoder,CodeLlama,PanGu-Coder} constructed upon the Transformer model architecture~\cite{Transformer}, achieving promising performance in code-related applications~\cite{wang2024tiger,wang2023boosting,wang2023generating,wang2024and,du2023classeval,zhang2024empirical,zhang2023malicious,du2024vul,yuan2023testchater,wang2024beyond}. These models are either pre-trained or fine-tuned on extensive code corpora, enabling them to automatically generate code based on provided natural language descriptions.
These code LLMs have demonstrated notable effectiveness in the generation of code blocks or functions. For instance, CodeLlama~\cite{CodeLlama}, built upon the foundational Llama2 model~\cite{Llama2}, has achieved state-of-the-art results among open code LLMs (\eg CodeGen~\cite{CodeGen} and StarCoder~\cite{StarCoder}), on benchmarks like HumanEval~\cite{Codex} and MBPP~\cite{corr/abs-2108-07732} that focus on standalone functions. 

\begin{figure}
    \centering
    \includegraphics[width=1\columnwidth]{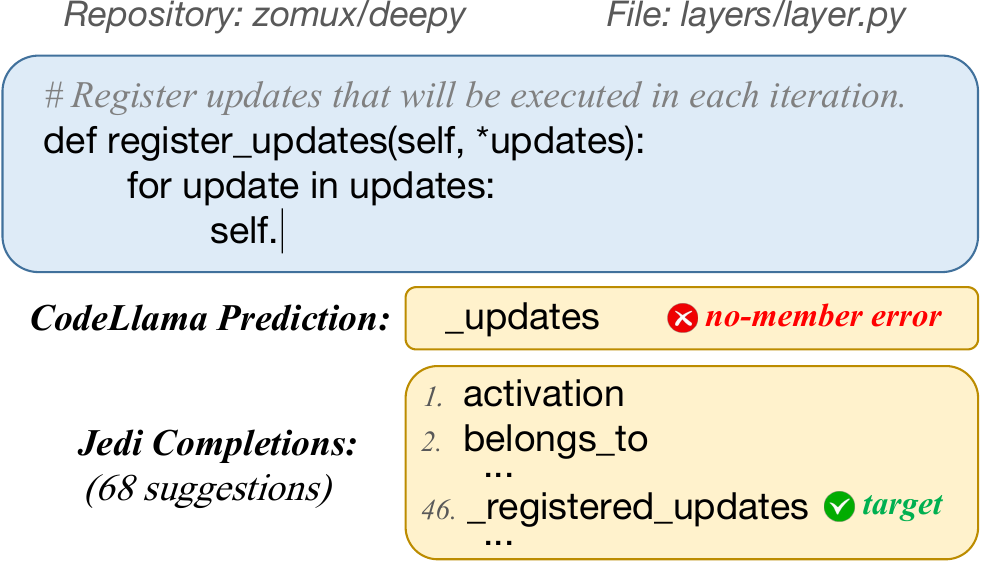}
    \caption{Illustrative Example of LLM Prediction and Tool Completion}
    \label{fig:motivation}
\end{figure}

% \textbf{Motivation.}
However, it is crucial to emphasize that in real-world code repositories, more than 70\% of functions are not standalone~\cite{yu2023codereval}. Code LLMs encounter significant challenges when generating such real-world functions, primarily because they cannot be aware of \textbf{\emph{repository-level dependencies}}, such as user-defined functions and attributes, during the code generation process~\cite{yu2023codereval}. This limitation often leads to the generation of code with \textbf{\emph{dependency errors}}, including \undefv and \nomem errors. These errors impede the usability and effectiveness of the code LLMs~\cite{Usability}.
For example, consider the scenario depicted in Figure~\ref{fig:motivation}. A code LLM (\eg CodeLlama) might incorrectly predict \inlinecode{\_updates} after generating \inlinecode{... self.}, resulting in a \nomem error because the object \inlinecode{self} does not possess an attribute named \inlinecode{\_updates}.

Meanwhile, modern Integrated Development Environments (IDEs) take a different approach, which typically incorporates code autocompletion tools based on program analysis. These tools, like Jedi~\cite{Jedi}, leverage their ability to analyze the current incomplete function's state and project context to provide \textbf{\emph{valid}} completion recommendations. This includes suggestions for accessible variables, attributes, and functions. 
For instance, when encountering \inlinecode{self.} in Figure~\ref{fig:motivation}, Jedi can infer and recommend 68 accessible attributes defined within \inlinecode{self}, including the target suggestion \inlinecode{\_registered\_updates}. Therefore, if we can seamlessly switch between code LLMs and the use of autocompletion tools, we have the potential to significantly reduce the occurrence of dependency errors in repository-level code generation.

% \textbf{Related Works.}
In fact, recent research has delved into the integration of external tools into the generation process of LLMs to mitigate their limitations in constrained generation scenarios. One noteworthy example is ToolFormer~\cite{Toolformer}, which creates an augmented dataset to instruct LLMs on invoking existing arithmetic calculators. This integration effectively reduces errors in generated text involving arithmetic calculations. Building upon ToolFormer's inspiration, Zhang \etal~\cite{zhang2023toolcoder} introduce ToolCoder, an approach designed to teach LLMs how to utilize \emph{information-retrieval-based} (IR-based) API search tools during the code generation process. While ToolCoder targets the generation of functionally correct standalone functions and demonstrates promising results, the integrated IR-based API search tools do not consider repository-level dependencies, limiting their potential in resolving dependency errors. \revise{Additionally, ToolFormer and ToolCoder are unable to handle scenarios where the tools return multiple candidates.}
Another relevant example of harnessing external tools is Repilot~\cite{wei2023copiloting}, which leverages code completion tools to filter out impractical suggestions made by LLMs in the context of automatic program repairing (APR). Unlike repository-level code generation, Repilot's primary focus is on generating valid \emph{single-hunk bug-fix patches} rather than entire functions. \revise{When applying Repilot to function-level code generation, the autocompletion tools are frequently triggered unnecessarily, resulting in significant overhead and impracticality.}

% \textbf{Challenges.}
In this work, we aim at integrating program-analysis-based code autocompletion tools into the generation process of code LLMs. Achieving the incorporation presents two key challenges. 
\textbf{\emph{(i) Determining when to trigger the invocation of autocompletion tools during the generation process:}} The generation process of LLMs is a step-by-step decoding process where each subsequent token is predicted based on previous tokens. In general, a function consists of dozens or even hundreds of tokens, making it impractical to invoke code autocompletion tools at every decoding step. In the case of tools like ToolFormer and ToolCoder, ChatGPT is employed to augment the training corpus by introducing special tokens into the text or code to mark positions where tool invocation is needed. After training on this augmented corpus, LLMs can predict the special token at the appropriate step, thereby triggering tool invocation. However, this ChatGPT-based augmentation method is less effective for repository-level code generation due to the presence of repository-level dependencies. The special token must be precisely inserted at positions involving such dependencies, such as when accessing user-defined variables.
\textbf{\emph{(ii) Selecting the target suggestion from the recommended completions of autocompletion tools:}} Different from tools like arithmetic calculation or API search integrated into ToolFormer and ToolCoder, which return a single result for each invocation, autocompletion tools often provide multiple completion suggestions sorted alphabetically. For instance, as depicted in Figure~\ref{fig:motivation}, Jedi returns a list of 68 completion suggestions (excluding builtin attributes), with the target suggestion being the 46th one in the list. Consequently, after invoking autocompletion tools, it is essential to assess the suggestions based on the generated code and select the most appropriate one. Furthermore, this selection process needs to be seamlessly integrated into the code generation process to ensure efficiency and coherence.

% \textbf{Approach.}
To tackle the challenges, we propose \App, an approach to integrate autocompletion tools into the generation process of code LLMs to support repository-level code generation. \App has two main phases: Trigger Insertion and Model Fine-tuning (Offline), and Tool-integrated Code Generation (Online). In the offline phase, \App analyzes source files within a corpus of code repositories, creating abstract syntax trees (ASTs) and extracting function definitions. It augments these functions by inserting a special token, \comp, signifying the positions to trigger autocompletion tools. The insertion positions are established by navigating through the functions and identifying the identifiers that can be recommended by autocomplete tools. These augmented functions, paired with their respective descriptions, are then employed to fine-tune a selected code LLM. In the online phase, \App iteratively constructs a function based on a provided description by predicting tokens step-by-step through the fine-tuned LLM. Whenever a \comp token is encountered, \App invokes the autocompletion tool to suggest code completions, drawing from the current repository context. Subsequently, it identifies the most appropriate suggestion through a constraint greedy search algorithm, appending this selected suggestion to the current tokens. This process continues as it predicts tokens until a specified termination condition is satisfied.

% \textbf{Evaluation.}
We conduct extensive experiments to evaluate the effectiveness of \App in repository-level code generation across three distinct code LLMs, namely, CodeGPT~\cite{CodeGPT}, CodeT5~\cite{CodeT5}, and CodeLlama~\cite{CodeLlama}. To facilitate this evaluation, we first construct a benchmark, which includes 12,406 Python functions from 671 real-world code repositories and 176 coding tasks from CoderEval dataset~\cite{yu2023codereval}. We define two new repository-level metrics, namely \cov and \valid. \cov quantifies the proportion of repository-level dependencies present in ground-truth functions and successfully covered by the generated functions, while \valid measures the percentage of generated functions that pass a dependency error check.
\revise{
The evaluation results on the 12,406 functions demonstrate that \App exhibits comparable or improved performance in widely-recognized similarity metrics such as BLEU-4, CodeBLEU, Edit Similarity, and Exact Match. Importantly, \App achieves significant improvements in \cov, ranging from 31.4\% to 39.1\%, and \valid, spanning from 44.9\% to 57.7\%, across the three code LLMs.
On the 176 tasks derived from CoderEval, \App achieves improvements of 40.0\% and 25.0\% in test pass rate (Pass@1) for CodeT5 and CodeLlama, respectively, while maintaining the same pass rate for CodeGPT. \App also demonstrates high efficiency in repository-level code generation, with average latency ranging from 0.63 to 2.34 seconds, attributed to offline fine-tuning with trigger insertion.}
Moreover, the results from our generalizability evaluation confirm that \App consistently performs well across a variety of code LLMs, with different model architectures and scales.
% The findings from the generalizability evaluation reaffirm the consistent performance of \App when applied to diverse code LLMs, encompassing various model architectures and scales.

% \textbf{Contributions.}
In summary, this paper presents the following key contributions:
\begin{itemize}
    \item \textbf{\App}, an novel approach that seamlessly integrates autocompletion tools into the generation process of code LLMs, which consists of Trigger Insertion and Model Fine-tuning (Offline), and Tool-integrated Code Generation (Online). \App seamlessly integrates the autocompletion tool into the generation process of code LLMs, thereby enhancing repository-level code generation. The offline phase results in an \textbf{Augmented Dataset}, which comprises 249,298 Python functions sourced from a diverse selection of 12,231 code repositories. Each function is augmented with a special token, \comp, which signifies positions suitable for invoking autocompletion tools.

    \item \textbf{An Evaluation Benchmark}, which encompasses 12,406 Python functions drawn from 671 real-world code repositories and 176 coding tasks with test cases derived from CoderEval, along with the introduction of two novel repository-level metrics: \cov and \valid.

    \item \textbf{Extensive Experimental Results}, which affirm the efficacy of \App in repository-level code generation. \App demonstrates substantial improvements in \cov, ranging from 31.4\% to 39.1\%, and \valid, spanning from 44.9\% to 57.7\%, across three distinct code LLMs. Additionally, \App achieves 40\% and 25\% improvements in test pass rate for CodeT5 and CodeLlama, respectively, with high generation efficiency.
\end{itemize}
\section{Preliminaries}\label{sec:definition}
% In this section, we formalize the tool-integrated generation process for repository-level code generation.

% \subsection{Preliminaries}

\subsection{Code LLMs}\label{sec:code-llm}
\begin{figure}
    \centering
    \begin{subfigure}{0.45\columnwidth}
        \centering
        \includegraphics[width=0.9\linewidth]{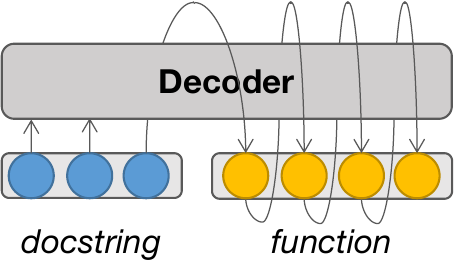}
        \caption{Decoder-only Model}
        \label{fig:decoder-only}
    \end{subfigure}
    \begin{subfigure}{0.45\columnwidth}
        \centering
        \includegraphics[width=0.9\linewidth]{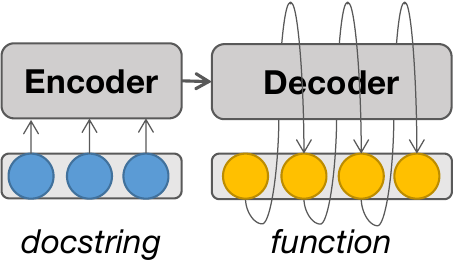}
        \caption{Encoder-Decoder Model}
        \label{fig:encoder-decoder}
    \end{subfigure}
    
    \caption{Decoder-only Model and Encoder-Decoder Model}
    \label{fig:models}
\end{figure}

Typically, there are two main categories of code LLMs that can be employed for code generation. These categories include decoder-only models and encoder-decoder models, each of which conducts the code generation process base on a given description as outlined below:
\begin{itemize}
    \item \textbf{Decoder-only Models:} Illustrated in Figure~\ref{fig:decoder-only}, decoder-only code LLMs, such as CodeGPT~\cite{CodeGPT} and CodeLlama~\cite{CodeLlama}, consist solely of a decoder component derived from the Transformer architecture~\cite{Transformer}. An employed decoder-only model first tokenizes the input description into a sequence of tokens. Subsequently, it feeds this token sequence into the model's decoder and proceeds to predict a function token-by-token, based on the context provided by the description and previously predicted tokens.

    \item \textbf{Encoder-Decoder Models:} As depicted in Figure~\ref{fig:encoder-decoder}, encoder-decoder code LLMs, such as CodeT5~\cite{CodeT5} and CodeT5+~\cite{CodeT5Plus}, encompass both the encoder and decoder components of the Transformer architecture. In this case, the employed model also tokenizes the description into a token sequence, but the sequence is first processed by the model's encoder. The model's decoder is then tasked with predicting a function token-by-token, relying on the representation produced by the encoder and the context provided by the preceding tokens.
\end{itemize}

On top of the standard generation process, to ensure that the employed code LLM can recognize and predict the special token \comp, we initially incorporate this token into the LLM's vocabulary, denoted as $\mathbb{V}_{llm}$. Formally, this addition results in an expanded vocabulary represented as:
\begin{equation}\label{eq:vocab}
    \mathbb{V} \leftarrow \mathbb{V}_{llm} \cup \{\comp\}
\end{equation}

For the employed code LLM, within the generation process, we define its tokenization process as a procedure:
\begin{equation}
    \textsc{llm-tokenize}: \Sigma_{char}^* \to \mathbb{V}^*
\end{equation}
Here, $\Sigma_{char}^*$ represents a character sequence of either a description or a code snippet, and $\mathbb{V}^*$ corresponds to the resulting sequence of tokens drawn from $\mathbb{V}$.

The next token prediction involved in each step is defined as a procedure:
\begin{equation}
    \textsc{llm-predict}: (\mathbb{V}^*, \mathbb{V}^*) \to [0,1]^{|\mathbb{V}|}
\end{equation}
In this context, the two input token sequences ($\mathbb{V}^*$) represent a description and an incomplete function, respectively, while $[0,1]^{|\mathbb{V}|}$ signifies a probability distribution encompassing $|\mathbb{V}|$ probabilities $[0,1]$. Here, $|\mathbb{V}|$ is the size (token numbers) of the vocabulary $\mathbb{V}$.

\textbf{\emph{Example:}} In Figure~\ref{fig:motivation}, CodeLlama takes the description ``Register updates...'' and the incomplete function ``... \texttt{ self.}'' as inputs. It then performs a prediction, generating a probability distribution of size $|\mathbb{V}|$, wherein the token \inlinecode{\_updates} exhibits the highest probability among all tokens.

\subsection{Autocompletion Tools}\label{sec:comp-tool}
An autocompletion tool takes a code repository and a caret position (defined as a tuple containing source file, line number, and column number) as input and provides a list of completion suggestions. We define this completion process as a procedure:
\begin{equation}
    \textsc{tool-complete}: (\Sigma_{repo}, \Sigma_{pos}) \to \Sigma_{iden}^{*}
\end{equation}
Here, $\Sigma_{repo}$ and $\Sigma_{pos}$ respectively represent the domains of code repositories and caret positions, $\Sigma_{iden}$ encompasses all possible identifiers such that $\Sigma_{iden}^{*}$ is a list of identifiers. It's worth noting that autocompletion tools often provide a wide range of completion suggestions, including keywords and partial identifiers. In our context, we focus solely on identifier-level completions, as keywords are relatively straightforward for code LLMs to predict, and partial identifiers are encompassed by identifier-level completions.

\textbf{\emph{Example:}} In Figure~\ref{fig:motivation}, when provided with the code repository and caret position, Jedi is capable of generating 86 completion suggestions for the incomplete function "... \texttt{ self.}".

\section{Approach}\label{sec:approach}
In this section, we elaborate on our approach named \App to integrate autocompletion tools into the generation process of code LLMs to support repository-level code generation.

\subsection{Overview}
\begin{figure*}
    \centering
    \includegraphics[width=2\columnwidth]{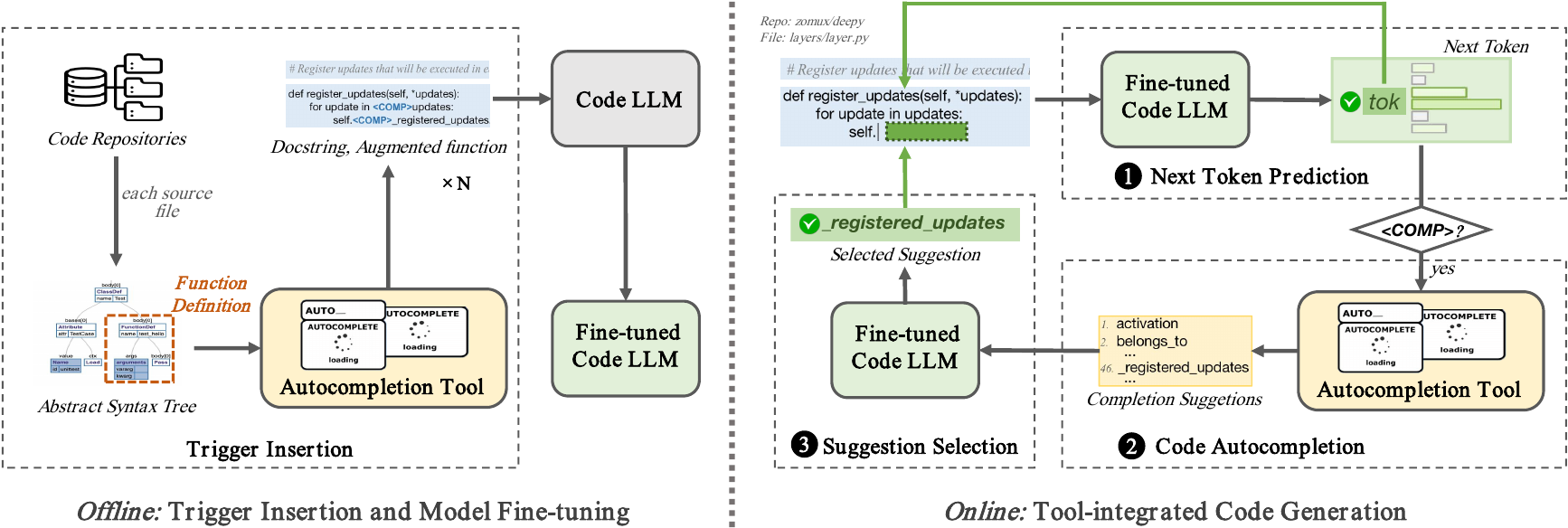}
    \caption{Approach Overview of \App}
    \label{fig:overview}
\end{figure*}

Figure~\ref{fig:overview} presents an overview of \App, which consists of two main phases, namely (i) \textbf{Trigger Insertion and Model Fine-tuning} (Offline) and (ii) \textbf{Tool-integrated Code Generation} (Online). 

In trigger insertion and model fine-tuning, \App parses each source file in the given code repositories into an abstract syntax tree (AST) and then extracts function definitions from the AST; For each extract function definition, \App then utilizes an autocompletion tool to augment it with the special token \comp to mark the positions to invoke the tool, and then assembles a pair of description and augmented function; After process all code repositories, \App employs the resulting pairs of descriptions and augmented functions to fine-tune a code LLM, resulting in a fine-tuned code LLM that can predict \comp at suitable positions to trigger the autocompletion tool.

In tool-integrated code generation, \App generates a token sequence to form a function by an iterative process, in which, at each step, one or multiple tokens are yielded by the fine-tuned code LLM and the employed autocompletion tool. At certain step, \ding{182} the fine-tuned code LLM takes the given description and the incomplete function as inputs and predicts the next token; The predicted token is appended to the incomplete function; \ding{183} If the predicted token equals \comp, the autocomplete tool is triggered and a list of completion suggestions is returned based on the current repository context;  \ding{184} \App then selects the most suitable one from the suggestions with the fine-tuned code LLM and appends the selected suggestion to the incomplete function.

\subsection{Trigger Insertion and Model Fine-tuning}
\subsubsection{Trigger Insertion}\label{sec:augmentation}
We employ a trigger insertion method to facilitate the learning process of code LLMs in determining when to utilize autocompletion tools during code generation. In this method, the special token \comp is inserted at specific locations within code functions, indicating when autocompletion tools should be triggered.

Given a code repository $\mathcal{R}$, we traverse each source file $file$ within it based on the file's suffix (\eg \texttt{.py} for Python) and then proceed to analyze the functions defined in the source file. To achieve this, we parse the source file into an abstract syntax tree (AST), where the functions are represented as function-definition nodes. Each function-definition node contains multiple AST-tokens, which are smallest individual units, such as keywords, identifiers, literals, operators, and punctuators, within programming language syntax. Note that these AST-tokens differ from the tokens in the LLM's vocabulary $\mathbb{V}_{llm}$. Typically, an AST-token comprises one or more tokens from $\mathbb{V}_{llm}$. For example, the AST-token \inlinecode{\_registered\_updates} consists of six tokens in vocabulary of CodeLlama, \ie [\inlinecode{\_}, \inlinecode{register}, \inlinecode{ed}, \inlinecode{\_}, \inlinecode{up}, \inlinecode{dates}].

For each function within the source file, we identify its corresponding function-definition node, denoted as $node$, and apply Algorithm~\ref{alg:parse_func} to it. The purpose of this algorithm is to traverse the function body and identify specific identifiers that are eligible for suggestions by autocomplete tools. Subsequently, the special token \comp is inserted in front of these chosen identifiers. More specifically, as the algorithm iterates through each AST-token $t$ within the function body $node.body$ (line 2), it performs two crucial checks. First, it employs the \textsc{isIdentifier} procedure to determine whether $t$ is an identifier. Second, it verifies that $t$ is not a built-in attribute, such as \inlinecode{\_\_dict\_\_} in Python, using the \textsc{isBuiltin} procedure. These conditions are essential because dependency errors often arise from user-defined attributes categorized as identifiers rather than other AST-tokens like language keywords. Additionally, these checks prevent the insertion of \comp at positions where the code LLM can confidently predict the following tokens, thus minimizing unnecessary tool invocations. When both conditions are met, the algorithm updates the caret position $\mathcal{P}$ to the start position of $t$ (line 4) and invokes the autocompletion tool to obtain a list of completion suggestions, denoted as $\mathbb{C}$ (line 5). If $\mathbb{C}$ contains $t$, indicating that the tool can propose the desired identifier, the special token \comp is inserted before $t$ to mark the position for triggering the autocompletion tool (lines 6-7). Upon executing the algorithm, we obtain the augmented function code $\mathcal{F}_{aug}$. 

\begin{algorithm}
\caption{Trigger Insertion}
\label{alg:parse_func}
\DontPrintSemicolon
    \KwInput{Repository $\mathcal{R}$, Source File $file$, Function-definition node $node$}
    \KwOutput{Augmented function $\mathcal{F}_{aug}$}
    % $node \leftarrow \textsc{getASTNode}(\mathcal{F})$ \\
    $\mathcal{F}_{aug} \leftarrow \textsc{getSignature}(node)$ \tcp*{signature}
    % \tcc{Traverse AST-tokens in the function body.}
    \For{$t$ \textbf{in} $node.body$} 
    {
        \If{$\textsc{isIdentifier}(t)$ \textbf{\emph{and not}} $\textsc{isBuiltin}(t)$}  {
            $\mathcal{P} = (file, t.start\_line, t.start\_column)$ \\
            $\mathbb{C} \leftarrow \textsc{tool-complete}(\mathcal{R}, \mathcal{P})$ \\
            \If{$t \in \mathbb{C}$} {
                $\mathcal{F}_{aug}.append(\comp)$
            }
            $\mathcal{F}_{aug}.append(t)$
        }
        
    }
\end{algorithm}

Next, we assemble a tuple $(\mathcal{D}, \mathcal{F}_{aug})$, in which $\mathcal{D}$ corresponds to the concatenation of the signature and docstring of the parsed function. Note that functions lacking corresponding docstrings are omitted from our process as our repository-level code generation relies on textual descriptions as input. Once we complete the processing of all code repositories, we accumulate an \textbf{\emph{augmented dataset}} that contains a substantial number of these data tuples.

\revise{
Note that our trigger insertion method can be applied to arbitrary code and is not limited to function bodies alone. Currently, we focus exclusively on function bodies, as our primary application scenario involves generating code based on the given natural language descriptions. Extracting descriptions for code blocks outside functions for model training and evaluation is challenging, due to the difficulty in determining the scope of line comments~\cite{ChenHLCZL19,HuangGDSCLZZ23}. Therefore, we solely consider function bodies, where corresponding descriptions can be readily obtained from function docstrings.
}

\begin{figure}
    \centering
    \includegraphics[width=1\columnwidth]{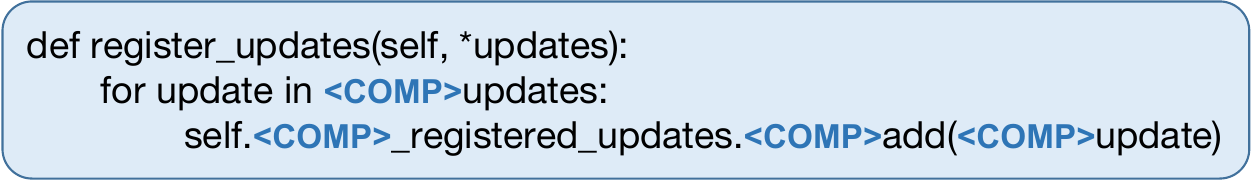}
    \caption{Augmented Function}
    \label{fig:aug_func}
\end{figure}

\textbf{\emph{Example:}}
In Figure~\ref{fig:aug_func}, we showcase an augmented function that contains four instances of the special token \comp. These tokens have been inserted at positions where the desired identifiers, namely \inlinecode{updates}, \inlinecode{\_registered\_updates}, \inlinecode{add}, and \inlinecode{update}, are found within the suggestion lists of the autocompletion tool.

\subsubsection{Model Fine-tuning}\label{sec:tuning}
During the fine-tuning process, we supply the collected descriptions and augmented functions to optimize the parameters of the employ code LLM (base model), adhering to established practices in code generation tasks. Specifically, for each pair consisting of a description $\mathcal{D}$ and an augmented function $\mathcal{F}_{aug}$, both are tokenized into sequences of tokens and subsequently fed into the base model to undergo the token-by-token generation process described in Section~\ref{sec:code-llm}. At each step, a cross-entropy loss is computed between the predicted probability distribution of the next token and the ground-truth next token present in $\mathcal{F}_{aug}$. 

In the case of code LLMs with an extensive number of parameters, such as CodeLlama-7B with 7 billion parameters, fine-tuning all parameters becomes computationally challenging due to resource limitations. To address this, we employ Low-Rank Adaptation (LoRA)~\cite{LoRA} as a parameter-efficient fine-tuning technique. LoRA relies on low-dimensional representations and a freeze-and-inject strategy, where the majority of the model parameters remain fixed, and trainable low-rank matrices are introduced into specific transformer layers, particularly the projection matrices within the attention module, to approximate weight updates.

\subsection{Tool-integrated Code Generation}
Based on the fine-tuned code LLM and the employed autocompletion tool, we perform a tool-integrated code generation process that is aware to the repository-level dependencies.

\subsubsection{Overall Process}
Algorithm~\ref{alg:generation} outlines the overall tool-integrated generation process, comprising three crucial parts based on the fine-tuned code LLM and the employed autocompletion tool: \ding{182} Next Token Prediction, \ding{183} Code Autocompletion, and \ding{184} Suggestion Selection. This algorithm takes a code repository $\mathcal{R}$, an insertion position $\mathcal{P}$, and a description $\mathcal{D}$ as inputs and follows an iterative process to generate a token sequence, ultimately constructing a function denoted as $\mathcal{F}$. Here, the tokens are drawn from the expanded vocabulary $\mathbb{V}$ defined in Equation~\ref{eq:vocab}. 

The iterative process commences with the \bos token (representing the beginning of the sequence), \ie $\mathcal{F} \leftarrow [\bos]$ in line 2, and proceeds by iteratively updating $\mathcal{F}$ until it reaches the \eos token (representing the end of the sequence). During each iteration step, the algorithm utilizes the description $\mathcal{D}$ and the current incomplete function $\mathcal{F}$ as inputs for the fine-tuned code LLM to execute the \textsc{llm-predict} procedure. This procedure predicts a $|\mathbb{V}|$-dimension probability distribution $\bm{p}^{|\mathbb{V}|}$ for the tokens in the vocabulary $\mathbb{V}$ (line 4). Subsequently, the token $tok$ with the highest probability is selected using the commonly used \textsc{argmax} function~\cite{Argmax} (line 5). The selected token $tok$ is then appended to $\mathcal{F}$ (line 6). If $tok$ corresponds to the \eos token, the iterative process concludes, yielding the final generated function (lines 7-8).

If $tok$ corresponds to the special token \comp
% and the probability $prob$ is greater than a predefined confidence threshold $CONF$
, the autocompletion tool is triggered to provide a list of completion suggestions denoted as $\mathbb{C}$. These suggestions are produced based on the code repository $\mathcal{R}$ and the caret position $\mathcal{P'}$ after inserting $\mathcal{F}$ at $\mathcal{P}$ (lines 9-11). Notably, when $\mathcal{F}$ is inserted using the $\textsc{insert}$ procedure, any \comp tokens within it are removed to prevent syntax errors. The fine-tuned code LLM is then employed to assess the completion suggestions and select the most suitable one for $\mathcal{F}$ by the \textbf{\textsc{llm-select}} procedure (line 12). The tokens from the selected suggestion are concatenated to $\mathcal{F}$.

\textbf{\emph{Example:}}
In the case of the incomplete code snippet shown in Figure~\ref{fig:motivation}, Algorithm~\ref{alg:generation} predicts the next token as \comp through the fine-tuned code LLM. This prediction triggers the autocompletion tool. Subsequently, the resulting completion suggestions are fed into the \textsc{llm-select} procedure, which determines the most appropriate suggestion.

\begin{algorithm}
\caption{Tool-integrated Code Generation}
\label{alg:generation}
\DontPrintSemicolon
    \KwInput{Repository $\mathcal{R}$, Description $\mathcal{D}$, Insertion Position $\mathcal{P}$}
    \KwOutput{Function $\mathcal{F}$}
    $\mathcal{D} \leftarrow \textsc{llm-tokenize}(\mathcal{D})$ \\
    $\mathcal{F} \leftarrow [\bos]$ \\
    \While{\textbf{true}} 
    {   
        \tcc{\ding{182} Next Token Prediction}
        $\bm{p}^{|\mathbb{V}|} \leftarrow \textsc{llm-predict}(\mathcal{D}, \mathcal{F})$ \\
        $tok \leftarrow \textsc{argmax}(\mathbb{V}, \bm{p}^{|\mathbb{V}|})$ \\
        $\mathcal{F} \leftarrow \mathcal{F} \oplus [tok]$ \\
        \If {$tok = \emph{\eos}$} {
            \textbf{break}
        }

        \tcc{\ding{183} Code Autocompletion}
        \If {$tok = \emph{\comp}$} {
        % \If {$tok = \emph{\comp}$ \textbf{and} $prob \geqslant CONF$} {
            % \tcc{suggest completions using tool}
            $\mathcal{P'} \leftarrow \textsc{insert}(\mathcal{P}, \mathcal{F})$ \\
            $\mathbb{C} \leftarrow \textsc{tool-complete}(\mathcal{R}, \mathcal{P'})$ \\
            % \tcc{select suitable one using LLM}
            \tcc{\ding{184} Suggestion Selection}
            $\textsc{\textbf{llm-select}}(\mathcal{D}, \mathcal{F}, \mathbb{C})$ \\
        }
        
    }
\end{algorithm}

\subsubsection{Completion Suggestion Selection}

\begin{figure}
    \centering
    \includegraphics[width=1\columnwidth]{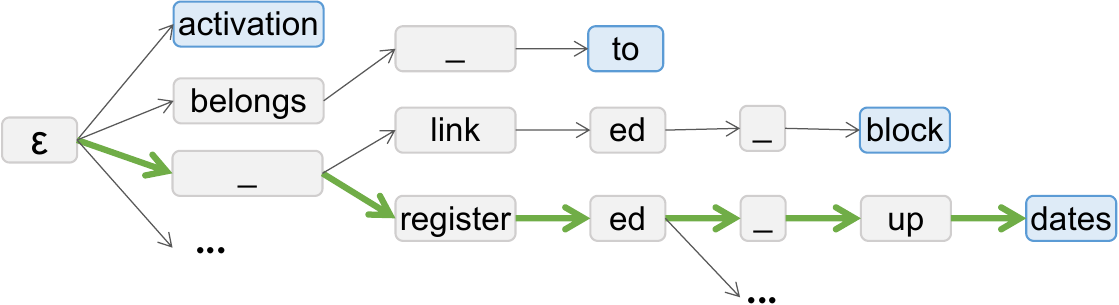}
    \caption{Example Prefix Tree}
    \label{fig:trie}
\end{figure}

Algorithm~\ref{alg:select} provides a description of the \textsc{llm-select} procedure, which is called within Algorithm~\ref{alg:generation}. To begin, it tokenizes each completion in $\mathbb{C}$ into a sequence of tokens from $\mathbb{V}$ using the code LLM's tokenizer (via the \textsc{llm-tokenize} procedure) and inserts this token sequence into a prefix tree~\cite{Trie}, denoted as $trie$ (lines 1-5). Each node in the tree possesses four properties: $node.token$, $node.tok\_idx$, $node.children$, and $node.is\_terminal$, indicating the token stored in the node, the index of the stored token in $\mathbb{V}$, the child nodes of the current node, and whether the node corresponding to the terminal of a token sequence. The root node, $trie.root$, is a unique node that stores $\epsilon$, signifying an empty string. Every path from $trie.root$ to a terminal node corresponds to a token sequence from $\mathbb{C}$. As an illustration, Figure~\ref{fig:trie} presents the prefix tree corresponding to the 68 completion suggestions shown in Figure~\ref{fig:motivation}. In this example, nodes enclosed in blue boxes indicate the terminals of token sequences.

\begin{algorithm}
\caption{Suggestion Selection based on Constraint Greedy Search}
\label{alg:select}
\DontPrintSemicolon
    \SetKwProg{Fn}{Procedure}{:}{}
    \Fn{\textsc{llm-select}{($\mathcal{D}, \mathcal{F}$, $\mathbb{C}$)}} {
        $trie \leftarrow Trie()$ \tcp*{prefix tree}
        \For{$comp$ \emph{\textbf{in}} $\mathbb{C}$} {
            $seq \leftarrow \textsc{llm-tokenize}(comp)$ \\
            $trie.insert(seq)$ \\
        }
        $node \leftarrow trie.root$ \\
        \While{\emph{\textbf{not}} $node.is\_terminal$} {
            $\bm{m}^{|\mathbb{V}|} \leftarrow \bm{0}$ \\
            \For{$child$ \emph{\textbf{in}} $node.children$} {
                $\bm{m}^{|\mathbb{V}|}[child.tok\_idx] \leftarrow 1$
            }
            $\bm{p}^{|\mathbb{V}|} \leftarrow \textsc{llm-predict}(\mathcal{D}, \mathcal{F})$ \\
            $\bm{p}^{|\mathbb{V}|} \leftarrow  \bm{p}^{|\mathbb{V}|} \odot \bm{m}^{|\mathbb{V}|}$ \\
            $tok \leftarrow \textsc{argmax}(\mathbb{V}, \bm{p}^{|\mathbb{V}|})$ \\
            $\mathcal{F} \leftarrow \mathcal{F} \oplus tok$ \\
            \For{$child$ \emph{\textbf{in}} $node.children$} {
                \If{$child.token = tok$} {
                    $node \leftarrow child$ \\
                    \textbf{break}
                }
            }
        }
    }
\end{algorithm}

Subsequently, the algorithm proceeds to select a path in $trie$ in a greedy fashion, based on predictions made by the fine-tuned code LLM, and appends the token sequence associated with the chosen path to the incomplete function $\mathcal{F}$ (lines 6-13). Specifically, the algorithm initiates a node pointer, denoted as $node$, with the root node $trie.root$ (line 6). A loop continues until the pointer $node$ reaches a terminal node (line 7). Within this loop, a $|\mathbb{V}|$-dimensional mask vector, denoted as $\bm{m}^{|\mathbb{V}|}$, is generated based on the children of the current node (lines 8-10). In $\bm{m}^{|\mathbb{V}|}$, only positions corresponding to the $tok\_idx$ property of the children of $node$ are assigned a value of 1, while all other positions are set to 0. Subsequently, the fine-tuned code LLM is employed to predict a probability distribution, $\bm{p}^{|\mathbb{V}|}$ (line 11). This predicted distribution is then element-wise multiplied by the mask vector $\bm{m}^{|\mathbb{V}|}$, effectively setting the probability of tokens not in the children of $node$ to 0. The next token, $tok$, is selected from $\mathbb{V}$ based on the highest probability in $\bm{p}^{|\mathbb{V}|}$ using the \textsc{argmax} function and is appended to the current incomplete function $\mathcal{F}$ (lines 13-14). Finally, the node pointer is updated to point to the child of $node$ whose stored token matches the selected token $tok$ (line 15-18).

\textbf{\emph{Example:}}
For the prefix tree illustrated in Figure~\ref{fig:trie}, the \textsc{llm-select} procedure iteratively selects the next tokens within the tree, guided by the LLM's predictions. This iterative process results in the inclusion of tokens corresponding to the suggestion \inlinecode{\_registered\_updates}, which are found along the \textcolor{LimeGreen}{\textbf{green path}}, being appended to the incomplete function.

\section{Evaluation Setup}

To evaluate the effectiveness and efficiency of \App in repository-level code generation, we conduct a comprehensive set of experiments.

\subsection{Research Questions}
We formulate the following research questions to guide our evaluation:

\begin{itemize}
    % \item \textbf{RQ1 - Impact of Hyperparameter:} What is the influence of the hyperparameter $CONF$ defined in Algorithm~\ref{alg:generation} on the generated code?

        % \item \textbf{RQ2 - Repository-level Metrics:} 
    
    \item \textbf{RQ1 - Similarity-based Effectiveness:} How closely does the code generated by \App align with the ground truth when assessed using common similarity metrics?

    \item \textbf{RQ2 - Dependency-based Effectiveness:} To what degree can \App cover repository-level dependencies and reduce dependency errors, including those related to user-defined functions and attributes?
    % \begin{itemize}
    %     \item  \textbf{RQ1.b:} What extent does the generated code cover repository-level dependencies, including user-defined functions and attributes?
    
    %     \item \textbf{RQ1.c:} To what degree can \App reduce errors, such as \nomem and \undefv errors, arising from repository-level dependencies?
    % \end{itemize}

    \item \revise{\textbf{RQ3 - Execution-based Effectiveness:} How effectively can \App generate functionally correct functions that pass test cases?}

    \item \revise{\textbf{RQ4 - Efficiency:} What is the average time \App takes to generate functions?}
    
    \item \textbf{RQ5 - Generalizability:} Is \App effective in code generation when applied to different code LLMs?
\end{itemize}

\subsection{Implementation}
Although \App is designed to be language-agnostic, our current focus is on developing a Python-specific prototype of \App.

\textbf{Base Model.}
In \App, we explore the utilization of three distinct code LLMs to encompass diverse model architectures and parameter scales. These code LLMs demonstrate impressive performance in code generation and have found extensive utilization in prior studies~\cite{CodeGPT,Compilable,CodeT5,CodeT5Plus,CodeLlama} for fine-tuning and evaluation.

\begin{itemize}
    \item \textbf{CodeGPT:} CodeGPT~\cite{CodeGPT} falls into the category of decoder-only models. It undergoes pre-training on a Python corpus sourced from the CodeSearchNet dataset~\cite{CodeSearchNet}, comprising 1.1 million Python functions. For our purposes, we adopt the pre-trained \emph{CodeGPT-small} version\footnote{https://huggingface.co/microsoft/CodeGPT-small-py}, which encompasses 124 million model parameters.

    \item \textbf{CodeT5:} CodeT5~\cite{CodeT5} belongs to the encoder-decoder model category and is similarly pre-trained on the Python corpus from the CodeSearchNet dataset. We select the pre-trained \emph{CodeT5-base} version\footnote{https://huggingface.co/Salesforce/codet5-base}, which comprises 220 million model parameters.

    % \item \revise{\textbf{CodeGen:} CodeGen~\cite{CodeT5} is also to an decoder-only model category and is trained on BigPython dataset that consists of 71.7B tokens of Python programming language. We select the pre-trained \emph{CodeGen-Mono-350M} version\footnote{https://huggingface.co/Salesforce/codegen-350M-mono}, which comprises 350 million model parameters.}

    \item \textbf{CodeLlama:} CodeLlama~\cite{CodeLlama} represents another decoder-only model, specialized for code-related tasks and based on Llama2~\cite{Llama2}. It is pre-trained on an even larger Python corpus, encompassing a staggering 100 billion tokens sourced from a Python-centric dataset~\cite{CodeLlama}. For our purposes, we adopt the pre-trained \emph{CodeLlama-7b} version\footnote{https://huggingface.co/codellama/CodeLlama-7b-Python-hf}, featuring a substantial 7 billion model parameters.
\end{itemize}
We refer to the variants of \App, namely \textbf{\AppGPT}, \textbf{\AppTFive}, and \textbf{\AppLlama}, corresponding to the underlying base models CodeGPT, CodeT5, and CodeLlama, respectively.
% Note that when feeding the code LLMs, we include the function signatures into the docstrings.

\textbf{Autocompletion Tool.}
We employ Jedi~\cite{Jedi} as our autocompletion tool. Jedi is a static analysis tool designed for Python, commonly utilized within integrated development environments (IDEs) and editor plugins. Utilizing Jedi, \App can trigger autocompletion, generating a list of suggestions that encompasses \emph{repository-level dependencies}, including user-defined attributes and functions.

\textbf{Trigger Insertion.}
To create the augmented dataset for fine-tuning the employed base model, we begin with the Python corpus from the \textit{training set} of CodeSearchNet dataset. Since the CodeSearchNet dataset does not provide complete code repositories from which to extract Python functions, we initiate the process by crawling the code repositories listed in the dataset. Subsequently, we follow the procedure outlined in Section~\ref{sec:augmentation} to extract and augment functions within these code repositories, ultimately generating the augmented dataset.
It's important to note that the CodeSearchNet dataset includes a partitioning into training, validation, and test sets. For our trigger insertion process, we exclusively utilize the code repositories associated with the training set. The resulting augmented dataset comprises a total of 249,298 pairs of descriptions and augmented functions, which are sourced from 12,231 distinct Python code repositories. Regarding dataset statistics, the average token count for descriptions is 10.98, and for augmented functions, it is 55.31. Additionally, the special token \comp appears an average of 5.54 times within these functions.

\textbf{Model Fine-tuning.}
In the fine-tuning process, we adopt different strategies for CodeGPT, CodeT5, and CodeLlama:
For CodeGPT and CodeT5, we perform full-parameter fine-tuning, optimizing all model parameters during this phase.
In the case of CodeLlama, we employ LoRA with a reduction factor (\emph{r}) of 8 and a scaling factor (\emph{alpha}) of 16 to achieve parameter-efficient fine-tuning. This approach allows us to optimize only 3.86\% of the trainable parameters in comparison to the original CodeLlama model.
The fine-tuning settings for learning rate and batch size are consistent across all three models, with a learning rate of 5E-6 and a batch size of 32. However, the number of epochs differs: 10 epochs for CodeGPT and CodeT5, while CodeLlama undergoes fine-tuning for 3 epochs. To ensure reproducibility, we set the seed for random functions to 42 consistently across all packages and libraries used.

% \textbf{Batched Generation Optimization.}
% \todo{details...}

\subsection{Evaluation Benchmark}
\subsubsection{Datasets}
To evaluate \App, we curate two datasets: (i) a large dataset derived from the CodeSearchNet~\cite{CodeSearchNet} to assess similarity-based and dependency-based effectiveness (RQ1 and RQ2); (ii) a dataset derived from CoderEval~\cite{yu2023codereval} containing test cases to evaluate execution-based effectiveness (RQ3).

\textbf{CodeSearchNet.} To assess similarity-based and dependency-based effectiveness, we follow this process to construct the dataset: We start by crawling the code repositories listed in the \textit{test set} of the CodeSearchNet dataset, ensuring no overlap with the \textit{training set} used for model fine-tuning. We then extract pairs of descriptions and functions from these repositories by parsing and traversing Abstract Syntax Trees (ASTs), similar to the method described in Section~\ref{sec:augmentation}. This process yields an \emph{evaluation dataset} comprising 12,406 Python functions sourced from 671 code repositories. On average, the descriptions contain 10.66 tokens, while the functions consist of an average of 54.54 tokens.

\revise{
\textbf{CoderEval.} To evaluate execution-based effectiveness, we initially gather all 230 Python code generation tasks from the CoderEval benchmark, extracted from 43 real-world Python repositories. Each task consists of a natural language description, a ground-truth code snippet, and a set of test cases, along with the project environment context associated with the task (e.g., project source code, dependent libraries, and test scripts). The tasks are categorized into six runnable levels: self-contained, slib-runnable, plib-runnable, class-runnable, file-runnable, and project-runnable~\cite{yu2023codereval}. Each runnable level relies on the dependencies defined at that level and does not depend on those defined at subsequent levels. For example, plib-runnable indicate that the task requires public third-party libraries, while file-runnable require dependencies defined in the current file (e.g., user-defined classes and functions). We remove the tasks overlapping with the training dataset of \App, resulting in a final dataset containing 176 tasks.
}

\subsubsection{Baselines}\label{sec:baselines}
The different variants of \App and the baselines are presented in Table~\ref{tab:baselines}, along with the base models they employ.

\textbf{Vanilla Baselines.}
We develop three vanilla baseline approaches by fine-tuning these same base models but performing straightforward code generation without tool integration. Specifically, the fine-tuning process for the baselines involves using the 249,298 pairs of descriptions and functions from the \emph{augmented dataset}. Notably, the fine-tuning is conducted on the original functions, prior to the introduction of \comp tokens. The training configurations, including learning rates and training epochs, mirror those employed in the implementation of \App. After fine-tuning, these models are utilized to perform straightforward code generation, as outlined in Section~\ref{sec:code-llm}. We label the three baseline approaches as follows:
\begin{itemize}
    \item \textbf{\BaseGPT}: Represents straightforward code generation using the CodeGPT model fine-tuned on original functions.
    \item \textbf{\BaseTFive}: Signifies straightforward code generation using the CodeT5 model fine-tuned on original functions.
    % \item \textbf{\BaseGen}: Signifies straightforward code generation using the CodeGen model fine-tuned on original functions.
    \item \textbf{\BaseLlama}: Designates straightforward code generation with the CodeLlama model fine-tuned on original functions.
\end{itemize}

\revise{
\textbf{Retrieval-Augmented-Generation (RAG) Baselines.}
We also include \Repo~\cite{zhang23repocoder}, a state-of-the-art approach that addresses repository-level code generation by integrating a similarity-based retriever and a pre-trained code language model in an iterative retrieval-augmented-generation pipeline. Similarly, we create three variants of \Repo based on the three fine-tuned models \BaseGPT, \BaseTFive, and \BaseLlama. We directly apply the prompt template defined in the original implementation of \Repo. The three variants of \Repo are listed as follows:
\begin{itemize}
    \item \textbf{\RepoGPT}: Represents the variant of \Repo with the CodeGPT model fine-tuned on original functions.
    \item \textbf{\RepoTFive}: Signifies the variant of \Repo with the CodeT5 model fine-tuned on original functions.
    % \item \textbf{\RepoGen}: Signifies the variant of \Repo with the CodeGen model fine-tuned on original functions.
    \item \textbf{\RepoLlama}: Designates the variant of \Repo with the CodeLlama model fine-tuned on original functions.
\end{itemize}
The hyperparameters of \Repo used in our experiments follow its default implementation: the retrieval-generation iteration is set to 2, the window size is 20, and the slice size is 2.
}

\revise{
\textbf{RAG-based Variants of \App.}
In fact, RAG method is orthogonal to our tool-integrated approach. To ensure a fair comparison and further explore the potential of \App, we also develop three RAG-based variants: \textbf{\RAGAppGPT}, \textbf{\RAGAppTFive}, and \textbf{\RAGAppLlama}. In these variants, the employed retrieval process is the same as the \Repo baselines.
}

\begin{table}
\caption{Variants of \App and baselines.}
\label{tab:baselines}
\setlength{\tabcolsep}{2pt}
\centering
\renewcommand{\arraystretch}{1.5}
    \begin{tabularx}{1\columnwidth}{lcccc} 
    \Xhline{2\arrayrulewidth}
    \textbf{Approach}                  &                      & \textbf{Base Model}         & \textbf{Architecture}               & \textbf{\# Parameters}           \\ 
    \Xhline{1.5\arrayrulewidth}
    \BaseGPT                           & \multicolumn{1}{c}{} & \multirow{4}{*}{CodeGPT}    & \multirow{4}{*}{Decoder-Only}       & \multirow{4}{*}{124 Million}   \\
    \RepoGPT                           & \multicolumn{1}{c}{} &                             & & \\
    \AppGPT (ours)                     & \multicolumn{1}{c}{} &                             & & \\
    \RAGAppGPT (ours)                  & \multicolumn{1}{c}{} &                             & & \\
    \hline
    \BaseTFive                         & \multicolumn{1}{c}{} & \multirow{4}{*}{CodeT5}     & \multirow{4}{*}{Encoder-Decoder}    & \multirow{4}{*}{220 Million}   \\
    \RepoTFive                         & \multicolumn{1}{c}{} &                             & & \\
    \AppTFive (ours)                   & \multicolumn{1}{c}{} &                             & & \\
    \RAGAppTFive (ours)                & \multicolumn{1}{c}{} &                             & & \\
    % \hline
    % \BaseGen                           & \multicolumn{1}{c}{} & \multirow{4}{*}{CodeGen}    & \multirow{4}{*}{Decoder-Only}       & \multirow{4}{*}{350 Million}   \\
    % \RepoGen                           & \multicolumn{1}{c}{} &                             & & \\
    % \AppGen (ours)                     & \multicolumn{1}{c}{} &                             & & \\ 
    % \RAGAppGen (ours)                  & \multicolumn{1}{c}{} &                             & & \\ 
    \hline
    \BaseLlama                         & \multicolumn{1}{c}{} & \multirow{4}{*}{CodeLlama}  & \multirow{4}{*}{Decoder-Only}       & \multirow{4}{*}{7 Billion}   \\
    \RepoLlama                         & \multicolumn{1}{c}{} &                             & & \\
    \AppLlama (ours)                   & \multicolumn{1}{c}{} &                             & & \\
    \RAGAppLlama (ours)                & \multicolumn{1}{c}{} &                             & & \\
    \Xhline{2\arrayrulewidth}
    \end{tabularx}
\end{table}

% This comprehensive set of baseline models allows us to assess and compare the performance of \App against straightforward code generation using the respective base models.

\subsubsection{Metrics}\label{sec:metrics}
In our evaluation, we employ commonly used similarity-based metrics, two novel dependency-based metrics, and an execution-based metric to evaluate the effectiveness of \App in repository-level code generation.

\textbf{Similarity-based Metrics:} We utilize the following well-established similarity metrics to measure the correspondence between generated functions and their ground-truth counterparts:
\begin{itemize}
    \item \textbf{BLEU-4}~\cite{BLEU}: This metric assesses the quality of generated code by comparing n-grams (sequences of n consecutive tokens) in the generated functions with those in the ground-truth functions.
    \item \textbf{CodeBLEU}~\cite{CodeGPT}: Specifically designed for code generation tasks, CodeBLEU evaluates the accuracy of code generation models by considering code-specific vocabulary and structure.
    \item \textbf{Edit Similarity (EditSim)}~\cite{EditSim}: This metric measures the similarity between two pieces of functions by analyzing the character-level edit operations required to transform one into the other.
    \item \revise{\textbf{Exact Match}: This metric measures the ratio of the generated code that are exactly matched with the ground truth.}
\end{itemize}

\noindent \revise{The calculation of the similarity-based metrics follows the implementation in CodeXGLUE\footnote{https://github.com/microsoft/CodeXGLUE}.}

\textbf{Dependency-based Metrics:} To assess the effectiveness of both \App and the baselines in repository-level code generation, we introduce the dependency-based metrics, namely \cov (DepCov) and \valid (ErrRate).
\begin{itemize}
    \item \textbf{\cov (DepCov):} This metric calculates the ratio of repository-level dependencies, including user-defined functions and attributes, that appear in ground-truth functions and are covered by the generated functions. 
    \revise{Given the $i$-th ground-truth function $gt_i$, we identify dependencies by performing the Trigger Insertion procedure (Algorithm~\ref{alg:parse_func}) and extracting expressions (such as function calls and attribute accesses like \inlinecode{self.\_registered\_updates}) that are marked with a trigger. These expressions are considered dependencies as their definitions can be traced in the current repository using static analysis tools like Jedi. Next, for the generated function $pred_i$ corresponding to $gt_i$, we extract all expressions by traversing its corresponding AST. We denote the identified dependencies in $gt_i$ and extracted expressions in $pred_i$ as two sets, $DEP(gt_i)$ and $EXP(pred_i)$, respectively. The \cov can be calculated as follows, where $N$ is the size of the test dataset:}
    $$DepCov = \frac{\sum_{i}^{N} |EXP(pred_i) \cap DEP(gt_i)|}{\sum_{i}^{N} |DEP(gt_i)|}$$

    \item \textbf{\valid (ValRate):} As repository-level dependencies can potentially introduce dependency errors in generated functions, we introduce the \valid metric (ValRate) to evaluate the effectiveness of \App in reducing dependency errors. This metric evaluates the proportion of generated functions that successfully pass a static check for dependency errors, specifically \nomem and \undefv. To perform this evaluation, we incorporate the generated functions into their respective code repositories and conduct static lint analysis using pylint~\cite{pylint}. Functions that do not exhibit syntax errors, \nomem, or \undefv errors are deemed statically valid. The \valid can be calculated as follows:
    $$ValRate = \frac{|\{pred_{i\leq N}: pred_i~\text{passes lint check}\}|}{N}$$
\end{itemize}

\revise{
\textbf{Execution-based Metric:} To further assess the functional correctness of the generated functions, we also employ a widely used execution-based metric involving running test cases.
\begin{itemize}
    \item \textbf{Test Pass Rate (Pass@1):} This metric calculates the ratio of generated functionally-correct functions that pass all corresponding test cases. It is evaluated specifically on the CoderEval dataset, where test cases and test scripts are provided.
\end{itemize}
}

% By incorporating these metrics, we aim to provide a comprehensive evaluation of the quality, coverage, and reliability of the generated code in the context of repository-level code generation.

\section{Results and Analyses}

\subsection{RQ1: Similarity-based Effectiveness}\label{sec:rq1-results}
The evaluation results of similarity-based metrics are presented in Table~\ref{tab:similarity}. When comparing the performance of \App's variants and the three different base models, namely CodeGPT, CodeT5, and CodeLlama, we observe that \App achieves similarity scores comparable to the baselines.

\begin{table*}
    \caption{Evaluation results of similarity-based effectiveness. $\Delta$ indicates the metric improvement or reduction of \App's variants compared to the baselines.}
    \label{tab:similarity}
    \centering
    \renewcommand{\arraystretch}{1.5}
    \begin{tabularx}{1.6\columnwidth}{lYYYY} 
    \Xhline{2\arrayrulewidth}
    \multirow{2}{*}{\textbf{Approach}}  & \multicolumn{4}{c}{\textbf{Similarity-based Metrics}}  \\
    \cline{2-5}
                                        &  BLEU-4      &       CodeBLEU      & EditSim    & ExactMatch                          \\ 
    % \cline{1-1}\cline{3-3}\cline{5-7}\cline{9-10}
    \Xhline{1.5\arrayrulewidth}
    \BaseGPT                            & 0.331  & 0.313    & 65.4\%   &  4.2\%                              \\
    \AppGPT                             & \makecell{0.340\\{\scriptsize ($\Delta=+2.7\%$)}}   & \makecell{0.310 \\{\scriptsize ($\Delta=-1.0\%$)}}     & \makecell{64.7\% \\{\scriptsize ($\Delta=-1.1\%$)}}       &         \makecell{4.7\% \\{\scriptsize ($\Delta=+11.9\%$)}}               \\ 
    \hline
    \BaseTFive                         & 0.341  & 0.289    & 63.9\%   &   4.3\%                             \\
    \AppTFive                          & \makecell{0.362 \\{\scriptsize ($\Delta=+6.2\%$)}}  & \makecell{0.293 \\{\scriptsize ($\Delta=+1.4\%$)}}    & \makecell{61.9\% \\{\scriptsize ($\Delta=-3.1\%$)}}     &         \makecell{5.5\% \\{\scriptsize ($\Delta=+27.9\%$)}}   \\ 
    % \hline
    % \BaseGen                           & 0.258  & 0.317    & 66.0\%                               \\
    % \AppGen                            & \makecell{0.267 \\{\scriptsize ($\Delta=+3.5\%$)}}  &\makecell{ 0.310 \\{\scriptsize ($\Delta=-2.2\%$)}}    & \makecell{65.3\% \\{\scriptsize ($\Delta=-1.1\%$)}}                             \\
    \hline
    \BaseLlama                         & 0.408  & 0.360    & 67.9\%     & 5.7\%                          \\
    \AppLlama                          & \makecell{0.425 \\{\scriptsize ($\Delta=+4.2\%$)}}  &\makecell{ 0.358 \\{\scriptsize ($\Delta=-0.6\%$)}}    & \makecell{66.3\% \\{\scriptsize ($\Delta=-2.4\%$)}}       &         \makecell{6.9\% \\{\scriptsize ($\Delta=+21.1\%$)}}                       \\
    \Xhline{2\arrayrulewidth}
    \end{tabularx}
\end{table*}

To provide a detailed breakdown, when utilizing CodeGPT as the base model, \AppGPT demonstrates a 2.7\% improvement in BLEU-4 and a 11.9\% improvement in Exact Match compared to \BaseGPT. However, it exhibits a 1.0\% decrease in CodeBLEU and a 1.1\% decrease in Edit Similarity. 
With the base model CodeT5, \AppTFive exhibits 6.2\%, 1.4\%, and 27.9\% enhancements in BLEU-4, CodeBLEU, and Exact Match relative to \BaseTFive but experiences a 3.1\% decrease in Edit Similarity. 
% For the base model CodeGen, \AppGen achieves a 3.5\% improvement in BLEU-4 compared to \AppGen but has a 2.2\% decrease in CodeBLEU and a 1.1\% decrease in Edit Similarity. 
In the case of the larger base model CodeLlama, \AppLlama shows improvements of 4.2\% and 21.1\% in BLEU-4 and Exact Match compared to \BaseTFive but encounters a 0.6\% decrease in CodeBLEU and a 2.4\% decrease in Edit Similarity.
% In summary, \App achieves improvements ranging from 0.8\% to 3.6\% in BLEU-4 but experiences decreases ranging from 1.0\% to 4.8\% in CodeBLEU and from 1.1\% to 5.0\% in Edit Similarity. 

% These results demonstrate that \App's performance remains competitive with the baselines across various base models, indicating the tool-integrated generation process does not cause significant loss of similarity.

\revise{
Although the absolute improvements in Exact Match rate are not large (from 0.5\% to 1.2\%), considering the size of the test set (\eg 12,406 samples), the additional exactly matched functions range from 62 to 149.
The variability in \App's performance across BLEU-4 and CodeBLEU can be attributed to the tokenization methods used in the calculations. For BLEU-4, before using the widely used utility script
\footnote{https://github.com/microsoft/CodeXGLUE/blob/main/Text-Code/text-to-code/evaluator/bleu.py}, we first tokenize the generated function and its corresponding ground-truth function using the tokenizer of the base code LLMs. In contrast, CodeBLEU is calculated based on the original generated and ground-truth code using the utility script\footnote{https://github.com/microsoft/CodeXGLUE/tree/main/Code-Code/code-to-code-trans/evaluator/CodeBLEU} that employs a simpler method, splitting functions into strings (\eg \inlinecode{func(arg1,)}) based on spaces. This splitting method may introduce inaccuracies in the statistics of matched n-grams, consequently affecting the CodeBLEU calculation.
For Edit Similarity, it is calculated at the character level, making it overly sensitive to semantics-insensitive elements like temporary variables. When two variables have different names, their similarity is much lower at the character level than at the token level.
}

\summary{
\App demonstrates competitive performance in similarity metrics compared to the baselines across various base models. It achieves improvements in BLEU-4 and Exact Match while exhibiting comparable performance in CodeBLEU and Edit Similarity.
}

\subsection{RQ2: Dependency-based Effectiveness}\label{sec:rq2-results}
\subsubsection{\cov}
Table~\ref{tab:dependency} displays the evaluation results for repository-level \cov (DepCov). Notably, our approach \App demonstrates significant superiority over the baselines across all three base models. Specifically, when employing the base models CodeGPT, CodeT5, and CodeLlama, \App surpasses the corresponding baselines in \cov by 39.1\%, 36.4\%, and 31.4\%, respectively.

\begin{table}
    \caption{Evaluation results of dependency-based effectiveness. DepCov and ValRate represent \cov and \valid, respectively. ValRate-\textit{dep} represents the \valid rate calculated only for functions containing dependencies. $\Delta$ indicates the metric improvement or reduction of \App's variants compared to the baselines.}
    \label{tab:dependency}
    \centering
    \renewcommand{\arraystretch}{1.5}
    \begin{tabularx}{1\columnwidth}{lYYY} 
    \Xhline{2\arrayrulewidth}
    \multirow{2}{*}{\textbf{Approach}}    & \multicolumn{3}{c}{\textbf{Dependency-based Metrics}}  \\
    \cline{2-4}
                                          &           DepCov     & ValRate   & ValRate-\textit{dep}                        \\ 
    \Xhline{1.5\arrayrulewidth}
    \BaseGPT                              &           8.7\%      & 50.4\%          &        46.5\%             \\
    \AppGPT                               &    \makecell{12.1\% \\{\scriptsize ($\Delta=+39.1\%$)}}  & \makecell{79.5\% \\{\scriptsize ($\Delta=+57.7\%$)}}  & \makecell{78.0\% \\{\scriptsize ($\Delta=+67.7\%$)}} \\ 
    \hline
    \BaseTFive                            &           11.0\%     & 47.3\%      &     42.5\%                     \\
    \AppTFive                             &    \makecell{15.0\% \\{\scriptsize ($\Delta=+36.4\%$)}}  & \makecell{70.6\% \\{\scriptsize ($\Delta=+49.3\%$)}}  & \makecell{68.0\% \\{\scriptsize ($\Delta=+60.0\%$)}} \\ 
    % \hline
    % \BaseGen                              &           10.0\%     & 51.6\%                               \\
    % \AppGen                               &    \makecell{13.7\% \\{\scriptsize ($\Delta=+37.0\%$)}}  & \makecell{71.7\% \\{\scriptsize ($\Delta=+39.0\%$)}} \\
    \hline
    \BaseLlama                            &           14.0\%     & 49.7\%       &    44.4\%                       \\
    \AppLlama                             &    \makecell{18.4\% \\{\scriptsize ($\Delta=+31.4\%$)}}  & \makecell{72.0\% \\{\scriptsize ($\Delta=+44.9\%$)}}  & \makecell{69.6\% \\{\scriptsize ($\Delta=+56.8\%$)}} \\
    \Xhline{2\arrayrulewidth}
    \end{tabularx}
    \vspace{-2mm}
\end{table}

These results underscore the effectiveness of the tool-integrated generation process in enhancing awareness of repository-level dependencies, a challenge often unaddressed by the conventional code LLMs. For instance, consider the incomplete function in Figure~\ref{fig:motivation}: in a straightforward CodeLlama generation, it fails to recognize the valid attributes of \inlinecode{self}. However, through tool-integrated generation, \App leverages Jedi to deduce a list of completion suggestions, enabling it to select the most suitable one and cover target repository-level dependencies, including the usage of user-defined attributes.

Despite the considerable improvement in repository-level \cov facilitated by our approach, it is essential to acknowledge that the overall coverage remains limited. This limitation arises from the fact that code LLMs generate function tokens sequentially from left to right. Consequently, errors tend to accumulate as the token count increases due to the exposure bias problem~\cite{BengioVJS15,PaulusXS18,AroraABC22}. This means that code LLMs often make incorrect token predictions at certain generation steps and may not produce \comp tokens to trigger autocompletion tools, especially for long functions.

\summary{
Our approach, \App, consistently outperforms the baselines in repository-level \cov across all three base models by ranging from 31.4\% to 39.1\%. These results highlight the effectiveness of our tool-integrated generation process in addressing the crucial issue of enhancing awareness of repository-level dependencies, which is often a challenge for conventional code LLMs in repository-level code generation.
}

\subsubsection{\valid}
Table~\ref{tab:dependency} also presents the evaluation results for \valid (ValRate and ValRate-\textit{dep}) in repository-level lint analysis, with a particular focus on \nomem and \undefv errors. Remarkably, our approach, \App, consistently exhibits significantly higher \valid compared to the baselines across all three base models. Specifically, when employing the base models CodeGPT, CodeT5, and CodeLlama, \App increases the \valid (ValRate) by 57.7\%, 49.3\%, and 44.9\%, respectively. \revise{When considering only the functions containing dependencies, \App improves the \valid  (ValRate-\textit{dep}) by 67.7\%, 60.0\%, and 56.8\%, respectively.}

These results underscore the effectiveness of our tool-integrated generation process in mitigating the production of invalid identifiers during code generation within a specific repository context. For instance, let's revisit the incomplete function in Figure~\ref{fig:motivation}: in a straightforward CodeLlama generation, it may predict a non-existent attribute, such as \inlinecode{updates}, for \inlinecode{self}. In contrast, through our tool-integrated approach, only valid completion suggestions inferred by Jedi are considered as candidates, thereby preventing numerous \nomem and \undefv errors.

\summary{
Our approach, \App, consistently achieves significantly higher \valid in repository-level lint analysis compared to the baselines, with improvements ranging from 44.9\% to 57.7\%. These results underscore the effectiveness of our tool-integrated generation process in mitigating the generation of invalid identifiers, a common challenge faced by conventional code LLMs in the context of repository-level code generation.
}

\subsection{RQ3: Execution-based Effectiveness}\label{sec:rq3-results}
\revise{
Table~\ref{tab:execution} presents the detailed evaluation results for test case execution (Pass@1) on the 176 CoderEval coding tasks.
}

\begin{table*}
    \caption{Evaluation results of execution-based effectiveness. \textit{self}, \textit{slib}, \textit{plib}, \textit{class}, \textit{file}, and \textit{project} represent self-contained, slib-runnable, plib-runnable, class-runnable, file-runnable, and project-runnable, respectively. The numbers in brackets after each runnable level indicate the corresponding number of tasks, while the numbers in brackets after the rates indicate the number of generated functions that pass the test cases. In each base model group, the best results are highlighted in \colorbox{gray!25}{gray}, except when all results are the same.}
    \label{tab:execution}
    \centering
    \renewcommand{\arraystretch}{1.5}
    \setlength{\tabcolsep}{2pt}
    \begin{tabularx}{1.8\columnwidth}{lYYYYYYY} 
    \Xhline{2\arrayrulewidth}
    \multirow{2}{*}{\textbf{Approach}}  & \multicolumn{7}{c}{\textbf{Execution-based Metric (Pass@1)}}  \\
    \cline{2-8}
                                        & \textit{total} (176) & \textit{self} (26) & \textit{slib} (23) & \textit{plib} (15) & \textit{class} (49) & \textit{file} (51) & \textit{project} (12) \\
    \Xhline{1.5\arrayrulewidth}
    \BaseGPT                           & \best{3.4\% (6)}    & \best{7.7\% (2)}    & 4.3\% (1)           & 6.7\% (1)        & 2.0\% (1)        & 2.0\% (1)        & 0.0\% (0)   \\
    \AppGPT (ours)                     & \best{3.4\% (6)}    & 3.8\% (1)           & 4.3\% (1)           & 6.7\% (1)        & 2.0\% (1)        & \best{3.9\% (2)} & 0.0\% (0)   \\
    \hdashline
    \RepoGPT                           & 2.8\% (5)           & 0.0\% (0)           & 4.3\% (1)           & 6.7\% (1)        & \best{4.1\% (2)} & 2.0\% (1)        & 0.0\% (0)   \\
    \RAGAppGPT (ours)                  & 2.8\% (5)           & 0.0\% (0)           & 4.3\% (1)           & 6.7\% (1)        & \best{4.1\% (2)} & 2.0\% (1)        & 0.0\% (0)   \\
    \hline
    \BaseTFive                         & 4.0\% (7)           & 7.7\% (2)           & 4.3\% (1)           & \best{13.3\% (2)}& 4.1\% (2)        & 0.0\% (0)        & 0.0\% (0)   \\
    \AppTFive (ours)                   & \best{5.1\% (9)}    & \best{15.4\% (4)}   & \best{8.7\% (2)}    & 6.7\% (1)        & 4.1\% (2)        & 0.0\% (0)        & 0.0\% (0)   \\
    \hdashline
    \RepoTFive                         & 4.0\% (7)           & 7.7\% (2)           & 4.3\% (1)           & \best{13.3\% (2)}& 4.1\% (2)        & 0.0\% (0)        & 0.0\% (0)   \\
    \RAGAppTFive (ours)                & \best{5.1\% (9)}    & \best{15.4\% (4)}   & \best{8.7\% (2)}    & 6.7\% (1)        & 4.1\% (2)        & 0.0\% (0)        & 0.0\% (0)   \\
    % \hline
    % \BaseGen                           & 4.0\% (7)           & 15.4\% (4)          & \best{8.7\% (2)}    & \best{6.7\% (1)} & 0.0\% (0)        & 0.0\% (0)        & 0.0\% (0)   \\
    % \RepoGen                           & 6.2\% (11)          & 15.4\% (4)          & 4.3\% (1)           & 13.3\% (2)       & 6.1\% (3)        & 0.0\% (0)        & 8.3\% (1)   \\
    % \AppGen (ours)                     & \best{5.1\% (9)}    & 15.4\% (4)          & \best{8.7\% (2) }   & \best{6.7\% (1)} & 6.1\% (3)        & \best{0.0\% (0)} & 0.0\% (0)   \\
    % \hdashline
    % \RAGAppGen (ours)                  & 5.7\% (10)          & 15.4\% (4)          & 8.7\% (2)           & \best{6.7\% (1)} & 6.1\% (3)        & 0.0\% (0)        & \best{16.7\% (2)}  \\
    \hline
    \BaseLlama                         & 6.8\% (12)          & 19.2\% (5)          & 13.0\% (3)          & 13.3\% (2)       & 4.1\% (2)        & 0.0\% (0)        & 0.0\% (0)   \\
    \AppLlama (ours)                   & 8.5\% (15)          & 23.1\% (6)          & 13.0\% (3)          & \best{20.0\% (3)}& 4.1\% (2)        & \best{2.0\% (1)} & 0.0\% (0)   \\
    \hdashline
    \RepoLlama                         & \best{10.8\% (19)}  & 26.9\% (7)          & \best{17.4\% (4)}   & 0.0\% (0)        & \best{10.2\% (5)}& \best{2.0\% (1)} & 16.7\% (2)  \\
    \RAGAppLlama (ours)                & \best{10.8\% (19)}  & \best{34.6\% (9)}   & 8.7\% (2)           & 6.7\% (1)        & 6.1\% (3)        & 0.0\% (0)        & \best{33.3\% (4)}  \\
    \Xhline{2\arrayrulewidth}
    \end{tabularx}
\end{table*}

\revise{
\textbf{Comparison to \Base baselines.}
Compared to the three \Base baselines, our approach \App generates 0, 2, and 3 additional functionally-correct functions, resulting in 0\%, 40.0\%, and 25.0\% improvements in Pass@1, respectively. Specifically, \AppGPT improves the pass rate for \textit{file-runnable} tasks, while reducing pass rate for \textit{self-contained} tasks; \AppTFive improves pass rate for \textit{slib-runnable} \textit{file-runnable} tasks; \AppLlama improves pass rate for \textit{self-contained}, \textit{plib-runnable}, and \textit{class-runnable} tasks. These tasks require different runnable-level dependencies (such as local variables and user-defined functions) to achieve correct functionality in the code. The enhancements by \App underscore the effectiveness of integrating autocompletion tools to handle these dependencies.
}

\revise{
\textbf{Comparison to \Repo baselines.}
Overall, the \Repo baselines show unstable performance across different base models. Specifically, compared to their respective \Base baselines, \RepoGPT and \RepoTFive exhibit reductions or no improvement in test pass rates, while \RepoLlama shows significant improvement. When compared to \RepoGPT and \RepoTFive, our \AppGPT and \AppTFive show improvements in pass rates with 1 and 2 more functions passing the test cases, respectively. However, \AppLlama exhibits a lower pass rate than \RepoLlama (15 vs. 19).
These variations can be attributed to several factors: CodeGPT and CodeT5 have fewer parameters (124 million and 220 million) and stricter token number limitations (1,024 and 512), limiting their ability to process and understand retrieval-augmented prompts. In contrast, CodeLlama, with more model parameters (7 billion) and support for longer token sequences (16,384 tokens), allows \RepoLlama to achieve a higher pass rate than \BaseLlama and \AppLlama due to the benefits of RAG.
}

\revise{
\textbf{Integration with RAG.} 
When integrating \App with RAG, \RAGAppGPT and \RAGAppTFive do not show improvement, while \RAGAppLlama exhibits breakthroughs for \textit{project-runnable} tasks, with 4 more generated functions passing the test cases. However, the overall pass rate remains unchanged after integrating RAG, showing different advantages and disadvantages of RAG integration for various runnable-level dependencies.
}

\revise{
\summary{
Our approach, \App, outperforms or matches the three \Base baselines (\BaseGPT, \BaseTFive, and \BaseLlama) by generating 0, 2, and 3 more functionally correct functions, achieving 0\%, 40.0\%, and 25.0\% improvements in Pass@1, respectively. Compared to RAG-based \Repo baselines, both \App and \Repo have their own advantages and disadvantages for different base models and runnable-level dependencies. Additionally, our decoding-stage tool integration approach shows potential when combined with prompt-level RAG techniques for addressing certain types of dependencies.
}
}

\subsection{RQ4: Efficiency}\label{sec:rq4-results}
\begin{figure}
    \centering
    \includegraphics[width=1\columnwidth]{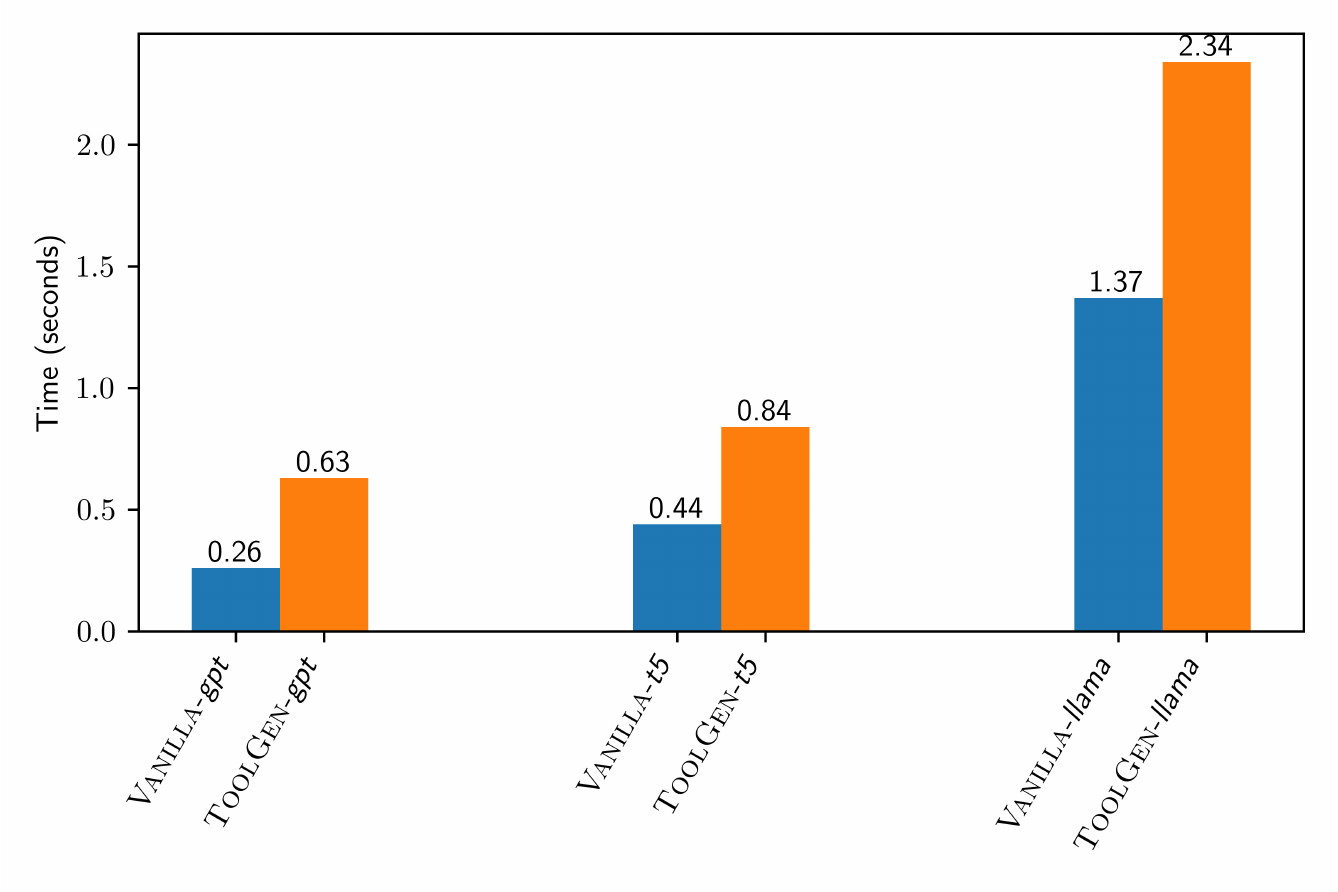}
    \caption{Results of efficiency evaluation.}
    \label{fig:time-plot}
\end{figure}

\revise{
Figure~\ref{fig:time-plot} illustrates the efficiency evaluation conducted on a single NVIDIA H100 Tensor Core GPU (80GB GPU Memory). Our approach exhibits approximately twice the average generation time for the 176 tasks in the CoderEval dataset, ranging from 0.63 to 2.34 seconds across different base models.
}

\revise{
The high efficiency of our tool integration is attributed to the offline trigger insertion and fine-tuning. The autocompletion tool is triggered only when the fine-tuned models predict the trigger token \comp, significantly reducing unnecessary tool invocations. Specifically, the fine-tuned CodeGPT, CodeT5, and CodeLlama predict an average of 5.02, 6.24, and 7.05 \comp tokens per task, respectively, which is much fewer than the average function length. Additionally, during the generation of a function, autocompletion is often triggered multiple times for the same objects (\eg \inlinecode{self}); we maintain a cache to recall completion suggestions for previously visited objects, thereby avoiding repeated tool invocations for the same objects.
}

\summary{
Our tool-integrated generation approach, \App, demonstrates high efficiency in repository-level code generation, with latency ranging from 0.63 to 2.34 seconds for generating each function. This efficiency is attributed to predicting the trigger token \comp and implementing a caching mechanism for completion suggestions.
}

\subsection{RQ5: Generalizability}\label{sec:rq5-results}
Based on the results presented in Table~\ref{tab:similarity} and Table~\ref{tab:dependency}, our tool-integrated generation approach consistently enhances performance in dependency-based metrics while maintaining comparable similarity-based metrics across various model architectures (decoder-only and encoder-decoder) and parameter scales (ranging from 124 million to 7 billion). According to Table~\ref{tab:execution} and Figure~\ref{fig:time-plot}, our approach improves or maintains execution-based metrics across the base models, with a consistent and acceptable additional latency overhead.

% Specifically, for the encoder-decoder model CodeT5, our tool-integrated generation approach demonstrates well-balanced and comprehensive enhancements in repository-level metrics. However, for the two decoder-only models, CodeGPT and CodeLlama, the improvements achieved by \App vary between \cov and \valid. When employing the smaller CodeGPT as the base model, \App significantly enhances \cov but exhibits less pronounced improvements in the \valid. In contrast, when employing the larger CodeLlama, the results are reversed. Upon closer examination of the generated functions, we attribute these variations to CodeGPT's occasional difficulty in predicting the \eos token correctly, leading to syntax errors that violate the prerequisites for successful lint check. In the case of CodeLlama, the initially high \cov, owing to its extensive parameter scale and training data, poses challenges for improvement through parameter-efficient fine-tuning (PEFT) with LoRA (only 3.86\% trainable parameters).

\summary{
Our tool-integrated generation approach consistently improves or maintains dependency-based and execution-based metrics while achieving competitive similarity-based metrics across various model architectures and parameter scales. This suggests that our approach is versatile and has the potential for broader applicability with other base models in repository-level code generation.
}

\section{Discussion}

\subsection{Case Study}\label{sec:case-study}
\begin{figure*}
    \includegraphics[width=2\columnwidth]{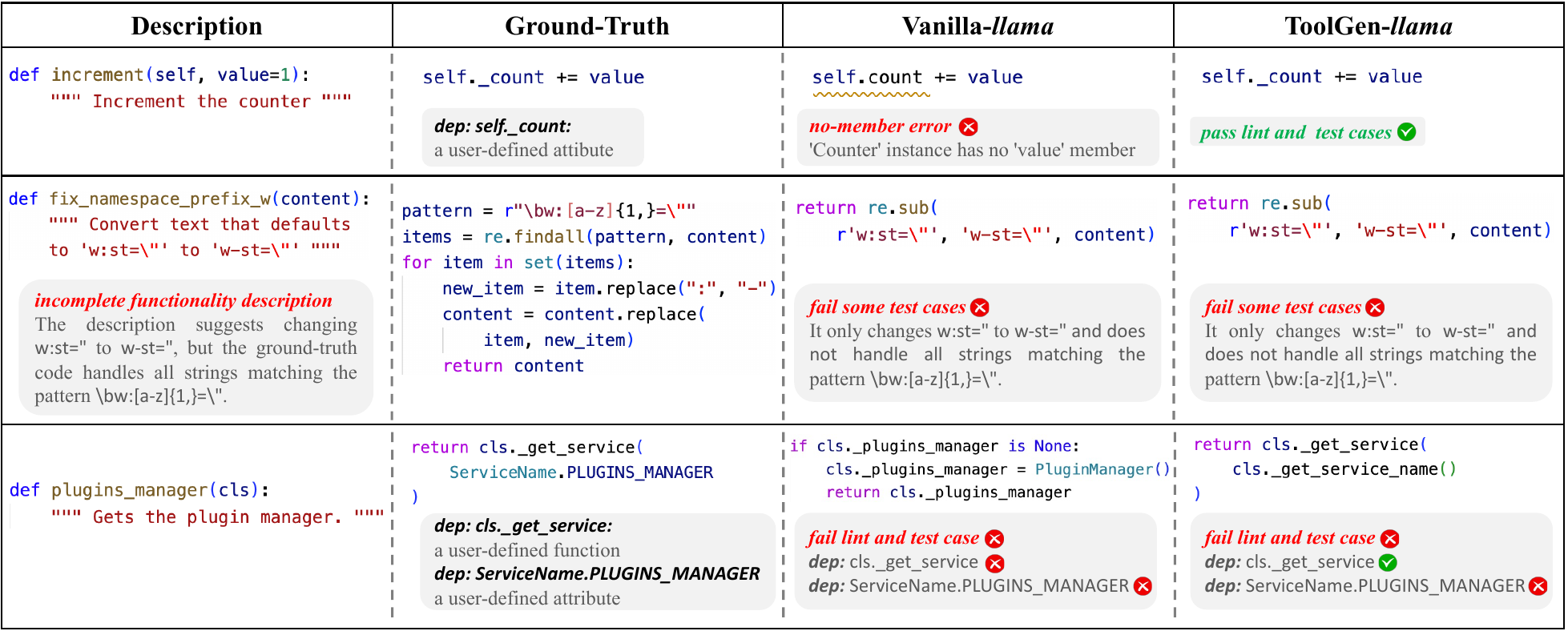}
    \caption{Case study of three specific examples. Additional explanatory notes are marked with \colorbox{gray!15}{gray boxes}. In the notes, "dep: xxx" denotes a dependency necessary in the generated code.}
    \label{fig:case-study}
\end{figure*}

\revise{
Figure~\ref{fig:case-study} depicts three specific examples using \BaseLlama and \AppLlama. Each row corresponds to an example, presenting the description, ground truth, code generated by \BaseLlama, and code generated by \AppLlama.
}

\revise{
\emph{Example 1:}  The code generated by \AppLlama successfully predicts the member \inlinecode{\_value} in the class \inlinecode{Counter}, while \BaseLlama incorrectly predicts an undefined member \inlinecode{value}, resulting in a \nomem error. This difference can be attributed to \App's integration of the autocompletion tool, which helps the code LLMs recognize necessary dependencies like user-defined attributes/members.
}

\revise{
\emph{Example 2:}  Both \BaseLlama and \AppLlama generate incorrect code that fails some test cases. After reviewing the description and the ground-truth, we find that the description is incomplete in expressing the desired functionality. As noted, the description only mentions changing \inlinecode{w:st="} to \inlinecode{w-st="}, but the actual desired functionality in the ground-truth is to handle all strings matching the pattern \inlinecode{\textbackslash bw:[a-z]\{1,\}=\textbackslash"}. Both \BaseLlama and \AppLlama follow the description and generate code that satisfies this incomplete functionality. This finding highlights the challenges posed by low-quality descriptions in real-world generation scenarios and reveals quality issues in existing benchmarks.
}

\revise{
\emph{Example 3:}  There are two crucial dependencies, namely \inlinecode{cls.\_get\_service()} and \inlinecode{ServiceName.PLUGINS\_MANAGER}, necessary to realize the required functionality. \BaseLlama fails to predict both dependencies and instead outputs non-existent dependencies like \inlinecode{cls.\_plugins\_manager} and \inlinecode{PluginManager()}, causing the generated code to fail lint checks and test cases. For \AppLlama, although it successfully predicts the dependency \inlinecode{cls.\_get\_service()}, it fails to predict \inlinecode{ServiceName.PLUGINS\_MANAGER} because the model chooses \inlinecode{cls} instead of \inlinecode{ServiceName} when starting predicting the argument for \inlinecode{cls.\_get\_service()}. This misleads the generation in an incorrect direction, resulting in the failure of the final code. This example also highlights the challenges of applying code LLMs in practical code generation, even when integrating autocompletion tools to avoid certain dependency issues. Introducing an incorrect token at any critical step in the generation process can result in the production of erroneous code.
}

\subsection{Limitations}\label{sec:limitations}

\emph{Static Autocompletion Tools for Dynamically Typed Programming Languages:}
Currently, our implementation of \App is specific to Python, a dynamically typed programming language. However, the autocompletion tools used in \App rely on static analysis, which can sometimes fail to trigger for certain repository-level dependencies. For instance, when the type of a function parameter cannot be explicitly inferred through static analysis, autocompletion tools may struggle to deduce attributes defined within the argument type. In the future, we plan to explore the integration of comprehensive type inference tools, such as learning-based tools, into the code generation process alongside autocompletion tools to enhance Python code generation.

\emph{Greedy Next Token Prediction in Generation Process:}
During the generation process, we employ a greedy strategy for next token prediction, where the token with the highest probability is selected using the \textsc{argmax} function. This greedy prediction strategy can occasionally lead the model to choose sub-optimal tokens for subsequent steps, resulting in code that may not be of the high quality. To address this issue, we intend to investigate the incorporation of techniques such as beam search and other advanced decoding methods into our tool-integrated generation process to mitigate the challenges posed by greedy prediction.

\emph{Dependency-based Evaluation Metrics:}
In the computation of the two repository-level evaluation metrics, namely \cov and \valid, we employ static analysis to identify target expressions and perform lint examinations. Similar to the autocompletion tools, these static tools may introduce a degree of inaccuracy into the calculated metrics. However, it is essential to note that this does not significantly impact the demonstrated effectiveness of \App, as the baseline metrics are also determined using the same static analysis.
% Another limitation is that our evaluation does not encompass metrics such as pass rate or pass@k for test cases, which are commonly used to assess the logical correctness of standalone functions generated by code LLMs. The primary reason for this omission is that such metrics typically rely on the execution of manually crafted unit test cases, which can be challenging to carry out in the context of repository-level code generation. Instead, our focus in this work centers on mitigating dependency errors, such as \nomem and \undefv errors, stemming from the unawareness of repository-level dependencies.

\revise{
\emph{Integration and Comparison with SOTA Closed-source LLMs:}  
Our approach can be applied to any encoder-decoder or decoder-only models. However, for the most state-of-the-art (SOTA) LLMs like GPT-3.5 and GPT-4, integrating the trigger insertion and tool-integrated decoding process faces challenges due to their closed-source nature and impracticality for fine-tuning. In the future, we may explore the possibility of integrating autocompletion tools into such closed-source LLMs in a non-tuning manner. 
In our evaluation, we do not compare \App with these SOTA closed-source LLMs, as our goal is to assess the effectiveness of integrating autocompletion tools for repository-level code generation. Therefore, we focus on comparing the performance of \App, \Base, and \Repo under the same base models.
}

\subsection{Threats to Validity}

\emph{Internal Threats:} The first internal threat pertains to potential data quality issues common in learning-based approaches. To mitigate this threat, we construct our augmented dataset and evaluation benchmark dataset using the widely adopted CodeSearchNet dataset, which serves as a reliable source for pretraining and evaluating various code models. Another internal threat pertains to the potential data leakage for CodeLlama, as the code repositories in the benchmark dataset may have been encountered by CodeLlama during its pretraining phase. However, our generalizability evaluation (RQ5) provides evidence of consistent performance across the three \App variants, suggesting that the improvements achieved by \AppLlama in repository-level code generation are not attributed to data leakage.
% The internal threat of randomness in model training, such as parameter initialization, may introduce reproducibility issues. To mitigate this risk, we fix random seeds at 42.

\emph{External Threats:} Our implementation and evaluation of \App are specific to the Python programming language. As a result, the findings may not be generalizable to other programming languages. Exploring the tool-integrated generation process for different languages is a valuable direction for future research.
% Another external concern relates to the applicability of \App to different code LLMs. To address this, we conduct an evaluation to assess the performance of \App across various model architectures and scales.
\section{Related Work}

\subsection{Code Generation}
Code generation has long been a central focus in the field of software engineering. Recent developments have introduced a range of large language models for code (code LLMs), including Codex~\cite{Codex}, CodeT5~\cite{CodeT5}, CodeT5+~\cite{CodeT5Plus}, InCoder~\cite{InCoder}, AlphaCode~\cite{AlphaCode}, CodeGen~\cite{CodeGen}, and CodeLlama~\cite{CodeLlama}, built upon the Transformer model architecture~\cite{Transformer}. These models, either pretrained or fine-tuned on extensive code corpora, have the capability to automatically generate code based on provided natural language descriptions.

While these code LLMs have demonstrated significant effectiveness in generating standalone functions on existing benchmarks like HumanEval~\cite{Codex} and MBPP~\cite{corr/abs-2108-07732}, they face substantial challenges when tasked with generating real-world functions within code repositories. The primary challenge stems from their lack of awareness of \emph{repository-level dependencies}, such as user-defined functions and attributes, during the code generation process~\cite{yu2023codereval}. To address these challenges, researchers have proposed prompt engineering approaches to make code LLMs aware of repository-level dependencies. Shrivastava \etal~\cite{ShrivastavaRepo} introduced the repository-level prompt generator, a framework for generating context-aware prompts without requiring access to the weights of the LLM. Bairi \etal~\cite{BairiRepo} presented CodePlan, a task-agnostic framework that treats repository-level coding as a planning problem, using innovative techniques to generate multi-step code edits while considering context from the entire codebase, previous changes, and specific instructions.

In this study, we tackle the challenges associated with repository-level code generation by seamlessly integrating autocompletion tools into the generation process of code LLMs.

\subsection{Incorporating External Tools into LLMs}
Recent research~\cite{Toolformer,zhang2023toolcoder,ART,Komeili0W22,LazaridouGSG22,Lamda,0001KARSW22,WebGPT,abs-2110-14168,PAL,POT} has explored the integration of external tools (\eg search engines, web browsers, calculators, and python interpreters) into the LLM generation process, aiming to address their limitations in certain generation scenarios. For instance, Schick \etal propose ToolFormer~\cite{Toolformer}, which augments datasets to instruct LLMs on invoking existing arithmetic calculators, effectively reducing errors in generated text related to arithmetic calculations. Building upon this idea, Zhang \etal introduce ToolCoder~\cite{zhang2023toolcoder}, designed to teach LLMs how to utilize information-retrieval-based (IR-based) API search tools during code generation. While ToolCoder is effective in generating functionally correct standalone functions, it falls short in addressing \emph{repository-level dependencies}, limiting its ability to resolve dependency errors. 
More relevant examples are Repilot~\cite{zhang23repocoder}, STALL+~\cite{liu2024stall}, and MGD~\cite{agrawal2023guiding}, which utilize code completion tools to filter out impractical suggestions made by LLMs, focusing on generating API/line-level code completions and valid bug-fix patches rather than entire functions.

In this paper, we integrate program-analysis-based autocompletion tools into the code LLM generation process to facilitate repository-level code generation.

\section{Conclusion}
We present \App, a novel approach that seamlessly integrates autocompletion tools into the code LLM generation process to effectively address repository-level dependencies. \App encompasses two crucial phases: Data Augmentation and Model Fine-tuning, and Tool-integrated Code Generation. Our comprehensive evaluation showcases \App's improvements in the two introduced dependency-level metrics and a widely used execution-based metric across three distinct code LLMs, while also demonstrating its competitiveness in widely-recognized similarity metrics. \App also demonstrates high efficiency in repository-level code generation, due to the offline fine-tuning with trigger insertion. Moreover, our generalizability evaluation reaffirms \App's consistent performance when applied to diverse code LLMs, including various model architectures and scales.

\bibliographystyle{IEEEtran}
\balance
\bibliography{src/citations.bib}

\end{document}